\documentclass[aps,preprintnumbers,amsmath,amssymb,nofootinbib,eqsecnum, preprintnumbers]{revtex4}
\usepackage{doi}
\usepackage{hyperref}
\hypersetup{
	colorlinks=true,        
	linkcolor=blue,         
	citecolor=cyan,         
}

\usepackage{eurosym}
\usepackage{amsfonts}
\usepackage{amsmath}
\usepackage{amssymb,epsf}
\usepackage{color}
\usepackage{graphicx}
\usepackage{natbib}
\usepackage{float}
\usepackage{caption}
\usepackage{subfig}
\usepackage{epstopdf}

\begin{document}
\title{Shadows and Optical Appearances of Black
Holes in $R^{2}$ Gravity}

\author{Khadije Jafarzade}
\email{khadije.jafarzade@gmail.com}
\affiliation{Department of Theoretical Physics, Faculty of Science, University of Mazandaran,
P. O. Box 47416-95447, Babolsar, IRAN}

\author{Sanjar Shaymatov}
\email{sanjar@astrin.uz}
\affiliation{Institute of Fundamental and Applied Research, National Research University TIIAME, Kori Niyoziy 39, Tashkent 100000, Uzbekistan}
\affiliation{Institute for Theoretical Physics and Cosmology, Zhejiang University of Technology, Hangzhou 310023, China}
\affiliation{University of Tashkent for Applied Sciences, Str. Gavhar 1, Tashkent 100149, Uzbekistan}
\affiliation{Western Caspian University, Baku AZ1001, Azerbaijan}

\author{Mubasher Jamil}
\email{mjamil@sns.nust.edu.pk}
\affiliation{School of Natural Sciences, National University of Sciences and Technology, Islamabad 44000, Pakistan}

\begin{abstract}

In this paper, we consider a charged AdS/dS black hole (BH) in $R^{2}$ gravity and study its optical features, including the shadow's geometrical shape and the energy emission rate. Additionally, we look for criteria to restrict the free parameters of the theory by comparing them to observational data of M87$^{\star}$. Then, we employ the Newman-Janis algorithm to build the rotating counterpart of the static solution in $R^{2}$  gravity and calculate the energy emission rate for the rotating case as well as discuss how the rotation factor and other parameters of this theory affect the emission of particles around the BHs. In the following, we consider the obtained rotating BH as a supermassive black hole and evaluate the parameters of the model with shadow size estimates based on the observations of M87$^{\star}$ from EHT.

\end{abstract}

\maketitle

\section{Introduction}
It is now well-understood that the Einstein's theory of General Relativity (GR), despite its many predictions which were found to be consistent with solar system and astronomical observations, is still an incomplete description of nature. GR breaks down at the spacetime singularities, do not prescribe the satisfactory explanation of the cosmological constant as well as the dark matter problem. Further GR is non-renormalizable and also incompatible with the principles of quantum mechanics. The problems mentioned above challenge GR and bring considerable interest in its extensions and alternative theories of gravity. Among the modified theories of gravity, quadratic gravity theory that extends the Einstein-Hilbert action by including the quadratic powers of the Ricci scalar and the Weyl tensor-stands out as an interesting theory \cite{Hell24JHEP}. $R^2$ Gravity, which only contains linear and quadratic terms in the Ricci scalar, is the simplest case of $f(R)$ gravity and the only ghost-free formulation within the quadratic gravity framework \cite{Sotiriou10Rev}. Unlike other models in the $f(R)$ family, $R^2$ gravity possesses a unique advantage-it is devoid of any inherent scale. This quadratic model is the simplest example of scale-invariant theory.  Adding other scale-invariant operators to the action, as in \cite{Salvio14JHE,Kannike15JHE}, the theory can be generalized to include the standard model fields,  playing a critical role in inflationary scenarios. Notably, recent studies have demonstrated that the inflationary phase of the early Universe is best described by $R^2$ gravity \cite{Sultana18Gen}. One intriguing feature of this theory is its ability to admit a non-trivial Ricci tensor $ R_{\mu\nu}\neq 0 $ even when $R=0$, in contrast to GR. $R^2$ gravity has a symmetry that is larger than scale symmetry and smaller than full Weyl symmetry,  the so-called restricted Weyl symmetry. This theory  is conformally equivalent to Einstein's gravity with a cosmological constant and a massless scalar field. The massless scalar can be identified as the Nambu-Goldstone boson arising from the spontaneously broken restricted Weyl symmetry \cite{Edery19PhysRev}. This equivalence ensures that $R^2$ gravity remains unitary, unlike other quadratic theories involving the square of the Ricci tensor, Riemann tensor, or Weyl tensor. The equivalence with Einstein gravity occurs when the restricted Weyl symmetry is spontaneously broken, which happens in the background (vacuum) spacetime with $ R\neq 0 $, including de Sitter (dS) and anti-de Sitter(AdS) spacetime but not Minkowski spacetime. In contrast, the symmetric $ R= 0 $ vacuum corresponds to the unbroken sector and has no correspondence to Einstein's gravity \cite{Edery18PhysRevD}. An important study on graviton propagators revealed that in a flat background, $R^2$ gravity propagates only a scalar mode, with no spin-2 field. However, in curved spacetime, the theory propagates both a massless spin-2 graviton and a spin-0 scalar state. Notably, the massless mode in curved spacetime enables long-range interactions, 
instead of a short-range Yukawa exchange \cite{Alvarez16Fortsch}.

If $R^2$ theory is formulated in the Einstein frame than the theory describes both gravity and a (either positive or negative choices for) cosmological constant resulting in dS or AdS spaces due to unbroken global scale symmetry \cite{Kehagias15JHEP}. In literature, the $R^2$ theory has been analyzed by its BH solutions and the resulting constraints from various solar system \cite{2024EPJC...84..330Z} as well as S2 star observations near the Milky Way galactic center \cite{2024JCAP...07..071Y}.

 As mentioned, there are unsolved challenges like the accelerated expansion of the universe. Although GR is the most accepted theory, it is not capable of explaining this completely. However, the accelerated expansion of the universe has been confirmed through observations regarding supernova explosions (SNIa). From this point of view, in GR field equations, the cosmological constant turns out to be a well-accepted candidate to provide a promising explanation for the accelerated expansion of the universe, acting as a vacuum energy $\Lambda$ along with a repulsive gravitational effect at large scales as well as close distances around the BH vicinity. Therefore, this has been considered dark energy in Einstein's field equation, leading to the accelerated expansion of the universe. Although the estimated magnitude of the cosmological constant $\Lambda$ as vacuum energy is of the order of $\sim 10^{-52}m^{-2}$, its repulsive gravitational nature significantly influences the accelerating rate of universe expansion at large enough distances~\cite{Peebles03,Spergel07}. The repulsive nature of the cosmological constant has been widely analyzed in many astrophysical situations since then \cite[see, e.g.,][]{Stuchlik05,Cruz05,Stuchlik11,Grenon10,Rezzolla03a,Arraut15,Faraoni15,Shaymatov18a,Giri23EPJP,Shaymatov21d,Rayimbaev-Shaymatov21a}. It must be noted that there exist other potential candidates for the accelerated expansion of the universe. Among them, the quintessence field can be considered a potential candidate for dark energy, explaining the expansion behavior of the universe. The quintessence field was proposed by Kiselev \cite{Kiselev03} and has also been considered a well-accepted alternative model \cite[see, e.g.,][]{Peebles03,Wetterich88,Caldwell09}.

Black holes (BHs) are among the most intriguing predictions of GR, with their existence confirmed by observational data. Investigating various aspects of black holes—such as their thermodynamic, optical, and dynamical properties—has significantly deepened our understanding of fundamental physics, astrophysics, and cosmology. Black hole thermodynamics, in particular, offers a remarkable framework for investigating the unification of GR and quantum mechanics. Hawking radiation, for instance, arises from quantum field theory applied in curved spacetime, highlighting the interplay between these two foundational theories. Understanding this relationship is essential for developing a complete theory of quantum gravity.  Black hole thermodynamics also encompasses intriguing phenomena such as phase transitions and thermal stability \cite{Jafarzade22Gen}, thermal fluctuations \cite{Rakhimova23Nuc}, heat engines \cite{Jafarzade20Nuc}, geometrical thermodynamics \cite{Jafarzade21Ann}, and the Joule-Thomson expansion \cite{Mustafa24JH,Javed24JHE}. Similarly, studying the optical and dynamic features of black holes—such as shadows and gravitational waves provides vital opportunities to distinguish a modified theory of gravity from GR and ensure the validity of the idea.


In a realistic scenario, it is widely believed that astrophysical BHs can be found immersed in non-vacuum matter fields that can, albeit small, affect both null and timelike geodesics and influence observable properties associated with the innermost stable circular orbits (ISCOs) \cite{Wald74,Shaymatov22a,Shaymatov21pdu}, quasi-periodic oscillations (QPOs) \cite{Shaymatov23ApJ,Mustafa24PDU1,Ditta24JHE,Shaymatov23EPJP,Mustafa24Phys,Shaymatov22c,Ashraf24Scr,Shaymatov20egb} and the size of the BH shadow \cite{Akiyama19L1,Akiyama19L6}. Recently, the Event Horizon Telescope (EHT) has detected plasma and polarized synchrotron radiation, providing a signature of magnetic fields near the M87$^{\star}$ supermassive BH \cite{EventHorizonTelescope:2021srq}. This discovery has motivated other researchers to investigate the effects of magnetized plasma effects on the BH shadow and accretion mechanisms \cite{Atamurotov:2021cgh,Pahlavon:2024caj,Azreg-Ainou:2024qqm}. The first observational image \cite{Akiyama19L1,Akiyama19L6} of the supermassive BH at the center of the host galaxy M87$^{\star}$ has definitively answered settled a long-standing fundamental question about the BH's image that had remained unanswered for decades, sparking increased research activity focused on analyzing the BH shadow.  Additionally, the EHT has recently detected an observational image of the supermassive BH in the host galaxy Sgr A$^{\star}$ with the discovery of a ring radius consistent with $10\%$ of GR prediction \cite{Akiyama22L12,Akiyama22L17}. This observation clearly demonstrates a bright ring-like behavior around the BH through the emitted light from the accretion disk. This is usually referred to as the BH image, showing a dark region in the sky surrounded by a light ring. Hence, the analysis of the accretion disk can provide promising insights into dark regions around BHs  \cite{Boshkayev:2020kle,Shaymatov2023,Boshkayev_2021,Alloqulov_2024}, including regions around AdS/dS BH in $R^{2}$. The first analysis regarding the dark region/disk was conducted for the Schwarzschild BH case \cite{Synge66}, providing an image of the BH shadow candidate \cite{Luminet79}. The analysis showing the formation of a BH shadow around a rotating Kaluza-Klein dilaton BH was theoretically conducted by \cite{Amarilla13} and then by \cite{Wei:2013kza} for the Einstein-Maxwell-Dilaton-Axion BH. This has led to active research focused on analyzing the shadow and light deflection effects to examine the spacetime geometry, particularly in the very close vicinity of the horizon. In recent years, within this research area, the shadow and light deflection analysis has been active in exploring the optical appearance of black holes from different perspectives \cite[see, e.g.,][]{Konoplya19PLB,Vagnozzi19PRD,Afrin21a,Atamurotov16EPJC,Konoplya19PRD,Atamurotov21JCAP,Mustafa22CPC,Tsukamoto18PRD,Rosa23PRD,Moffat20PRD,Al-Badawi24EPJCa,Al-Badawi24CPC,Jafarzade21JCAP,Jafarzade21PRD,Jafarzade23PTEP,Jafarzade24CQG,Al-Badawi24EPJCb}. 

It is to be emphasized that the EHT not only detected the shadow of the supermassive BHs at the center of the host galaxies M87$^{\star}$ \cite{Akiyama19L1,Akiyama19L6} and Sgr A$^{\star}$ \cite{Akiyama22L12,Akiyama22L17} but also helped provide unique insights into BHs. These triumphs associated with the shadows of these supermassive BHs through the EHT have opened a new avenue for excellent tests in probing the nature of BHs along with their surrounding geometries. Additionally, these EHT observations are playing a pivotal role in constraining the validity of alternative models to BHs and the parameters of BHs within different theories of gravity \cite{Kocherlakota21EHT,Kumar22ApJ,Walia22ApJ,Sengo23JCAP,Jafarzade22Ann,Hendi23,Jafarzade24PDU,Al-Badawi24CTP,Ashraf25PDU}. With this motivation, in this paper, we consider charged and rotating AdS/dS BHs in $R^{2}$ gravity and examine their optical appearances and energy emission rates. This analysis enhances our understanding of the unique insights into the nature of BHs in $R^{2}$. Additionally, we utilize EHT observations and compare their precise measurements with theoretical results to obtain the best-fit constraints on the BH parameters in $R^2$ gravity.

The paper is organized as follows: In Sec.~\ref{Sec:metric}, we briefly discuss charged BH solutions in $R^2$ gravity, along with the admissible parameter space between BH parameters beyond which no BH solution exists. In Sec.~\ref{Sec:shadow}, we study the optical properties of the charged BH, including the effects of the spacetime parameters arising from $R^2$ gravity. In Sec.~\ref{Sec:rot_bh}, we extend the static and spherically symmetric charged BH solution in $R^2$ gravity to a Kerr-like rotating charged BH solution by adapting the Newman-Janis algorithm (NJA), providing insights into its parameters through the analysis of the BH shadow and energy emission rate based on observations from the EHT. Finally, we end up with our remarks and conclusions in Sec.~\ref{Sec:conclusion}. We use $(–, +, +, +)$ for the spacetime metric and the system of units $c=G=1$ throughout the paper.

\section{\label{Sec:metric} Charged black hole solutions in $R^{2}$ gravity}
In this section, we review charged BH solutions in $R^{2}$ gravity and find the allowed regions to have physical BH solutions. The $R^2$ theory of gravity coupled to matter (electromagnetic field) is defined by the following
action \cite{Kehagias15JHEP}
\begin{eqnarray}\label{sa}
S=\int d^4 x\sqrt{-g}\left(\frac{1}{16 \mu^2}R^2-\frac{1}{4}F_{\mu\nu}F^{\mu\nu}\right)\, ,
\label{Se} 
\end{eqnarray}
in which $ R $ is the Ricci scalar and $ F_{\mu\nu}=\partial_{\mu}A_{\nu}- \partial_{\nu}A_{\mu}$ indicates the Faraday tensor ($ A_{\mu} $ is the gauge potential). Also, $ \mu $ denotes a non-zero cosmological constant. The equations of motions which follow from the variation of the action Eq.~(\ref{sa}) by the metric and the 4-vector potential, respectively are given by
\begin{eqnarray}
&&R\, R_{\mu\nu}-\frac{1}{4}R^2\, g_{\mu\nu}-\nabla_\mu\nabla_\nu R
+g_{\mu\nu}\nabla^2 R=4 \mu^2\left(F_{\mu\rho}{F_{\nu}}^\rho-\frac{1}{4}g_{\mu\nu} F^2\right)\, ,  \label{Eeq}\\
&&\nabla_\mu F^{\mu\nu}=0 \label{max}\, .
\end{eqnarray}

A static and spherically symmetric solution is simply written as follows:
\begin{eqnarray}
ds^2=-f(r)dr^2+\frac{dr^2}{f(r)}+r^2 d\Omega_2^2\, ,
\end{eqnarray}
together with the electric field resulting from the Maxwell equation 
\begin{eqnarray}
A_t=\frac{Q}{r}, ~~~F_{tr}=\frac{Q}{r^2}\, .
\end{eqnarray}
The field equations Eq.~(\ref{Eeq}) can be rewritten in terms of the metric function $f(r)$ as 
\begin{eqnarray}
 \Big{(}r f'+2 f\Big{)} \Big{[}r^4 f''^2+f \left(8-8 r
   f'\right)+4 r f' \left(r^2 f''+2\right)-4
   f^2+8\mu^2Q^2-4\Big{]}=0\, ,
 \end{eqnarray} 
 which solves to give  
 \begin{eqnarray}
 f(r)=1-\frac{M}{r}-\frac{Z}{r^2}+\frac{Q^2\mu^2}{6Z}r^2\, ,
 \label{static-solution}
 \end{eqnarray}
 where $M$ and $Q$ are BH mass and electric charge respectively, while $Z$ is a coupling parameter. Finally, the BH solution within this modified theory can then be rewritten as \cite{Kehagias15JHEP}
\begin{eqnarray}
&& d\bar s^2=-\left(1-\frac{M}{r}-\frac{Z}{r^2}+\frac{Q^2\mu^2}{6Z}r^2\right)dt^2+\frac{dr^2}{1-\frac{M}{r}-\frac{Z}{r^2}+\frac{Q^2\mu^2}{6Z}r^2}+r^2 d\Omega_2^2\label{sol1}\, ,\\
&&A_t=\frac{Q}{r}\, .
\end{eqnarray}

\begin{figure}[!htb]
\centering
\subfloat[]{
        \includegraphics[width=0.44\textwidth]{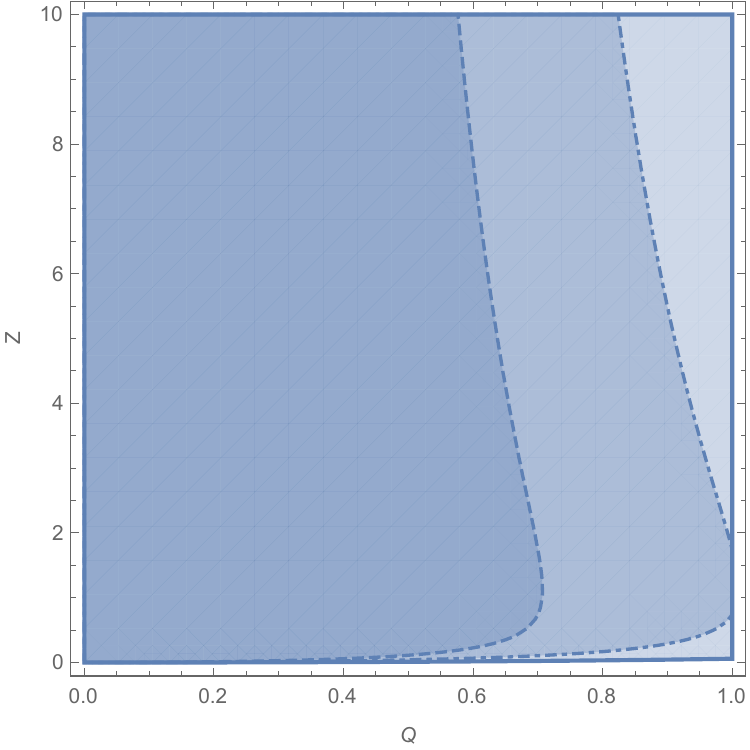}}
\subfloat[]{
     \includegraphics[width=0.46\textwidth]{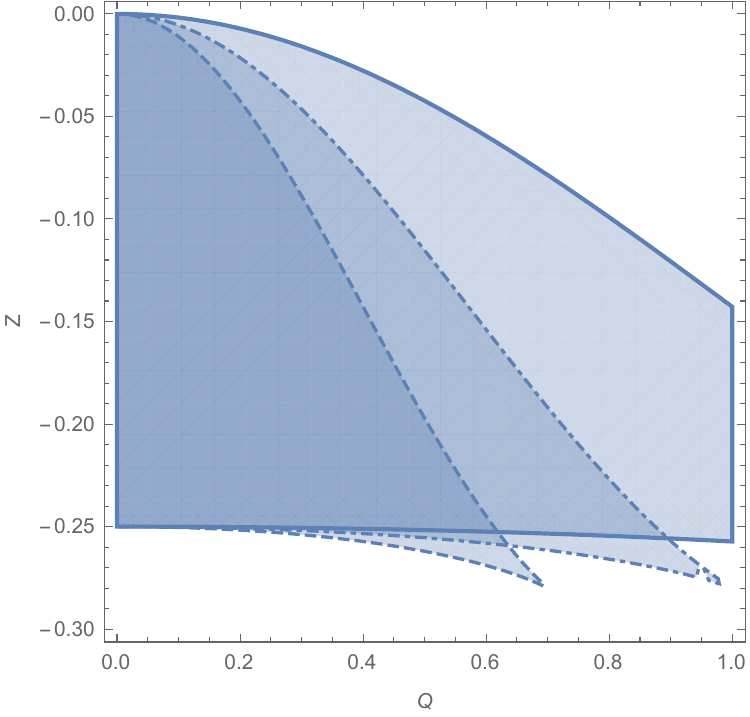}}
\caption{The admissible parameter space plot (shaded areas) between the parameters $Q$ and $Z$ for various combinations of cosmological constant $\mu=0.4,\, 0.7,\, 1$ corresponding to solid, dash-dotted, and dashed curves, respectively. Note that the white area refers to the region, where no black hole solution exists.}
\label{Fig1}
\end{figure}

From Eq.~(\ref{sol1}), the new term, $Q^2\mu^2/Z\neq 0$, is involved in the metric function, leading to non-asymptotically flat solutions. This results in the allowed solutions of de-Sitter Reissner-Nordstr\"om (dS-RN) for $Z<0$ and anti-de Sitter Ressner-Nordstrom (AdS-RN) type solutions for $Z>0$. Now, we are interested in examining the allowed regions of the parameters in which a physical BH solution can exist. In Fig.~\ref{Fig1}, we demonstrate the admissible parameter space plot ($Q-Z$) for various combinations of the cosmological constant $\mu$. As can be seen from Fig.~\ref{Fig1}, the admissible parameter space region decreases as the non-zero cosmological parameter $\mu$ increases in both $\pm\,Z$ cases. It is valuable to note that the BH region that can exist in the shaded region separated from no BH regions by curves is larger than the one for $Z<0$.

\section{\label{Sec:shadow} Charged BH's optical properties in $R^{2}$ gravity}

\subsection{The charged BH shadow }
Here, we now consider the general formalism used to model and analyze the motion of the photon in the close vicinity of the charged BH spacetime, as described by the line element given in Eq.~(\ref{static-solution}). For that we employ the null geodesic equations to determine the radius of innermost circular orbit for the photon around the BH spacetime considered here. The Hamiltonian of the system for the photon around a static and spherically symmetric BH spacetime is defined by  \cite{Hendi23}
\begin{eqnarray}
H=\frac{1}{2}g^{ij}p_{i}p_{j}=0\, ,  \label{EqHamiltonian}
\end{eqnarray}
where $ p_{i}=g_{ij}\dot{x}^{j} $ is the generalized momentum. Following to the spherically symmetric property of the BH spacetime, we further consider null geodesics of photons on the equatorial plane (i.e., $\theta =\pi/2$). Here, recalling the Hamiltonian of the system for the photon motion, we rewrite Eq.~(\ref{EqHamiltonian}) as follows: 
\begin{eqnarray}
\frac{1}{2}\left[ -\frac{p_{t}^{2}}{f(r)}+f(r)\, p_{r}^{2}+\frac{p_{\phi }^{2}}{%
r^{2}}\right] =0\, .  \label{EqNHa}
\end{eqnarray}

It is well-known fact that the Hamiltonian does not depend explicitly on the coordinates $t$ and $\phi$, thus leading to define the constants of motion, such as $p_{t}=-E$ and $p_{\phi}=L$. Note also that $E$ and $L$ respectively refer to the specific energy and angular momentum of the photon. Employing the Hamiltonian formalism, one can further obtain the equations of motion as
\begin{eqnarray}
\dot{t}=\frac{\partial H}{\partial p_{t}}=-\frac{p_{t}}{f(r)},~~~~~~\dot{r}%
=\frac{\partial H}{\partial p_{r}}=p_{r}f(r)\mbox{~~~and~~~}\dot{\phi}=\frac{%
\partial H}{\partial p_{\phi }}=\frac{p_{\phi }}{r^{2}}\, ,
\end{eqnarray}%
where the dot denotes the derivative with respect to the affine parameter and
$p_{r}$ is the radial momentum. Following to the equations of motion, together with two
conserved quantities, $E$ and $L$, we obtain the null geodesic equation 
\begin{eqnarray}
\dot{r}^{2}+V_{\mathrm{eff}}(r)=0\, ,  \label{EqVef1}
\end{eqnarray}%
where $V_{\mathrm{eff}}$ refers to the effective potential of the photon and is defined by 
\begin{eqnarray}
V_{\mathrm{eff}}(r)=f(r)\left[ \frac{L^{2}}{r^{2}}-\frac{E^{2}}{f(r)}\right]\, .  \label{Eqpotential}
\end{eqnarray}
\begin{figure}[!htb]
\centering
\subfloat[$ Z=0.1 $]{
        \includegraphics[width=0.32\textwidth]{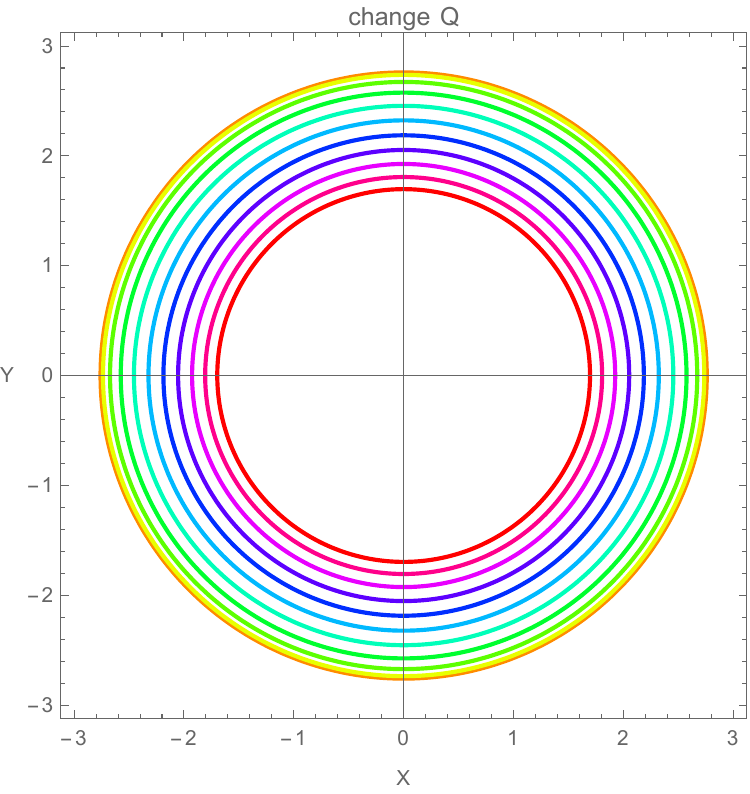}}
\subfloat[ $ Q =0.2 $ ]{
     \includegraphics[width=0.32\textwidth]{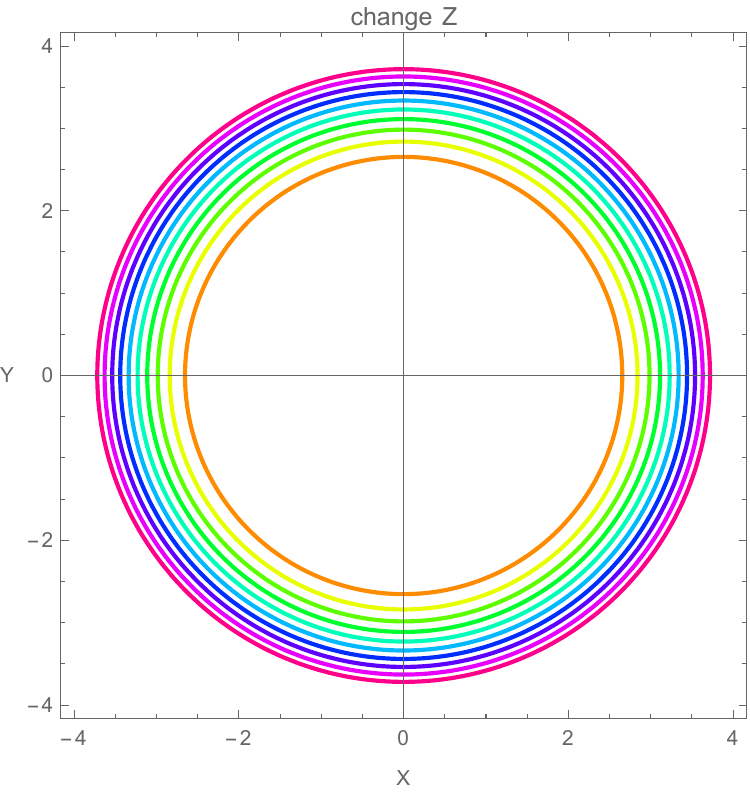}}
   \subfloat[$ Q =0.2 $]{
        \includegraphics[width=0.32\textwidth]{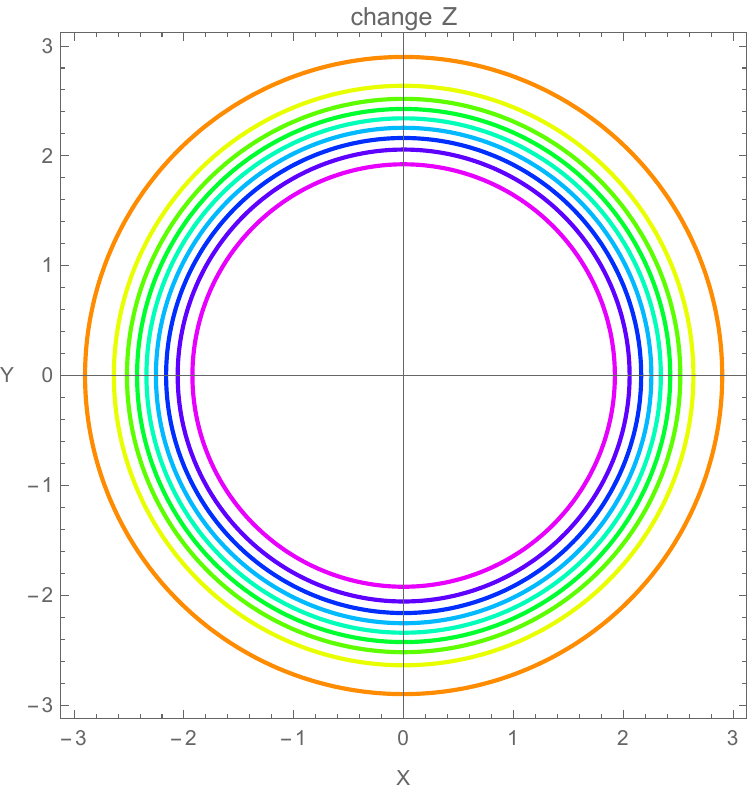}}      
\newline
\caption{\label{Fig2} The dependence of the boundary of BH shadow on the parameters $Q$ and $Z$ while keeping $M=1$ and $\mu=0.4$ fixed. Left panel: The boundary of the BH shadow is plotted for various values of the parameter $Q$ ranging from 0 to 0.8 with equal intervals, corresponding to the curves from outermost (orange) to the innermost (red). Middle panel: The boundary of the BH shadow is plotted for various values of the parameter $Z$ ranging from 0.1 to 1.0 with equal intervals, corresponding to the curves from innermost (orange) to the outermost (magenta). Right panel: The boundary of the BH shadow is plotted for various values of the parameter $Z$ ranging from -0.3 to -0.03 with equal intervals, corresponding to the curves from innermost (purple) to the outermost (orange). }
\end{figure}

We then further explicitly demonstrate the existence of circular null geodesics.  The conditions for circular null geodesics are given by 
\begin{eqnarray}
V_{\mathrm{eff}}(r_{ph})=0 \mbox{~~~and~~~} V_{\mathrm{eff}}^{\prime }(r_{ph})=0\, ,
\label{EqVeff1}
\end{eqnarray}%
where the former defines the critical angular momentum $L_{p}$ of the
photon sphere while the latter determines the radius of the photon sphere, i.e. $r_{ph}$, giving 
\begin{eqnarray}
    r_{ph}f'(r_{ph})-2f(r_{ph})=0. \label{phs}
\end{eqnarray}
By virtue of Eqs.~(\ref{static-solution}) and (\ref{phs}) and the shadow radius definition, we can determine the size of the BH shadow \cite{Perlick22}  
\begin{eqnarray}
r_{sh}=\frac{L_{p}}{E}=\frac{r_{ph}}{\sqrt{f(r_{ph})}}\, .  \label{Eqrsh}
\end{eqnarray}


We now move to analyze the BH shadow resulting from the BH charge, non-zero cosmological and coupling BH parameters within the modified theory of $R^2$ gravity. In Fig.~\ref{Fig2}, we demonstrate the boundary of the BH shadow around the charged BH in $R^2$ gravity for various combinations of parameters $Q$ and $Z$. In Fig.~\ref{Fig2}, the left panel shows the impact of the BH charge $Q$ on the shadow boundary, while the middle and right panels show the impact of coupling parameter $\pm\, Z$ for the fixed cosmological parameter $\mu$. As can be seen from Fig.~\ref{Fig2}, the charge $Q$ decreases the size of the BH shadow, resulting in the shape of shadow sphere shifting toward to smaller $r_{sh}$. Unlike $Q$, the BH shadow boundary/radius increases as a consequence of an increase in the coupling parameter $Z>0$ while keeping $Q$ and $\mu$ fixed. However, it should be noted that the BH shadow decreases with an increasing the coupling parameter $Z<0$, similarly to what is observed in the BH shadow boundary resulting from the charge parameter $Q$.    

\subsection{Observational constraints from the EHT observations}
\label{IIIB}
\begin{figure}[!htb]
\centering
\subfloat[]{
        \includegraphics[width=0.44\textwidth]{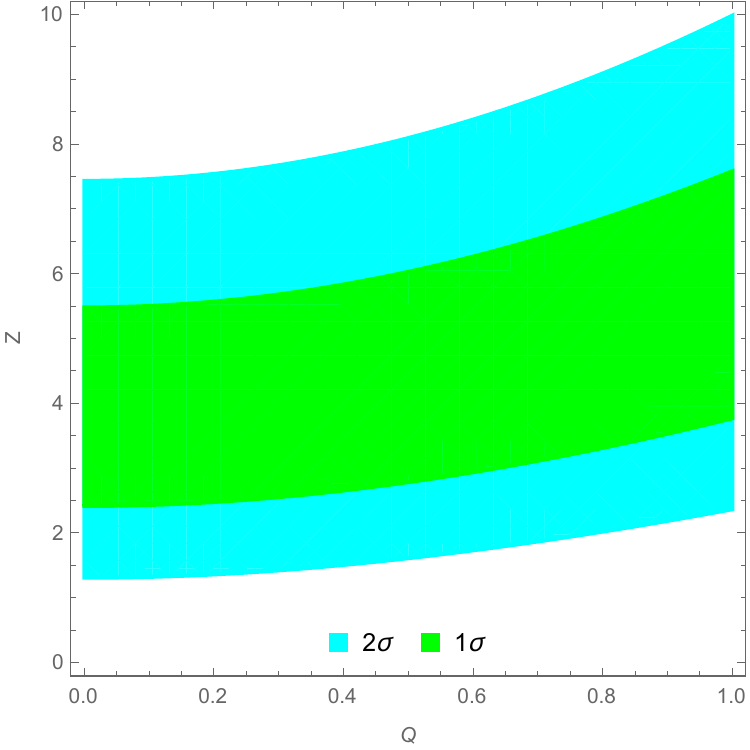}}
\subfloat[]{
     \includegraphics[width=0.46\textwidth]{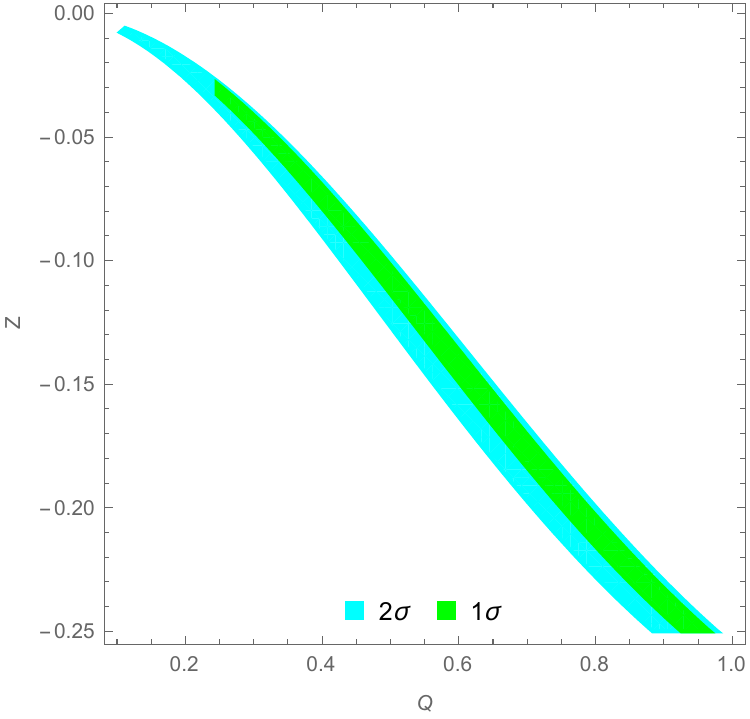}}
     \newline
\caption{Constraints on the $R^{2}$ gravity's parameter and the charge within the EHT observations of M87$^{\star}$ for keeping the cosmological constant $\mu=0.4$ fixed. Note that we set $M=1$. }
\label{Fig3}
\end{figure}
\begin{figure}[!htb]
\centering
\subfloat[]{
        \includegraphics[width=0.44\textwidth]{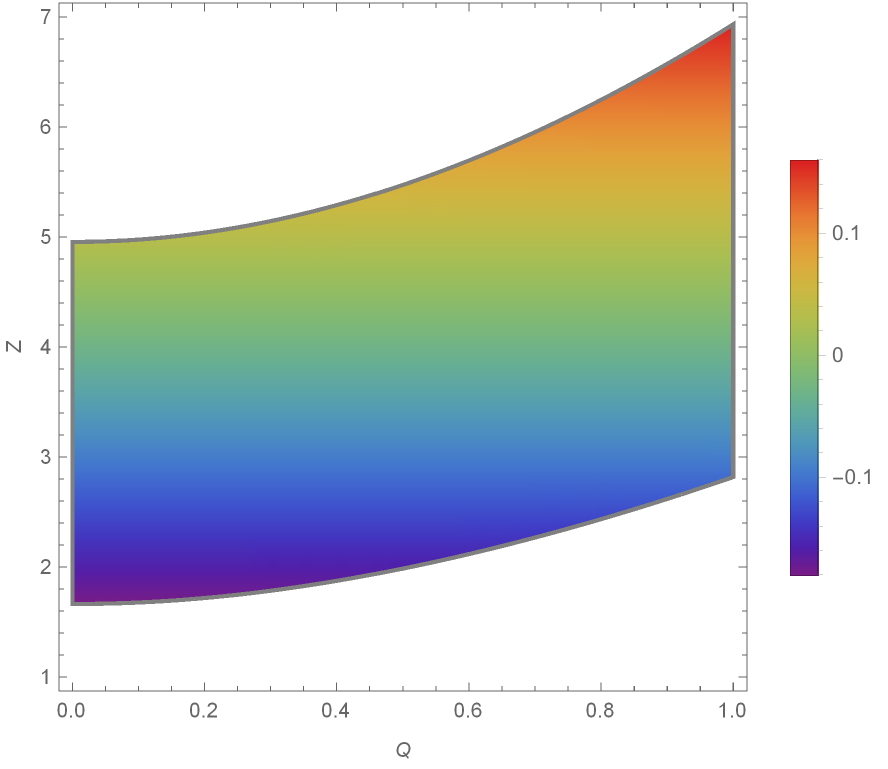}}
\subfloat[]{
     \includegraphics[width=0.46\textwidth]{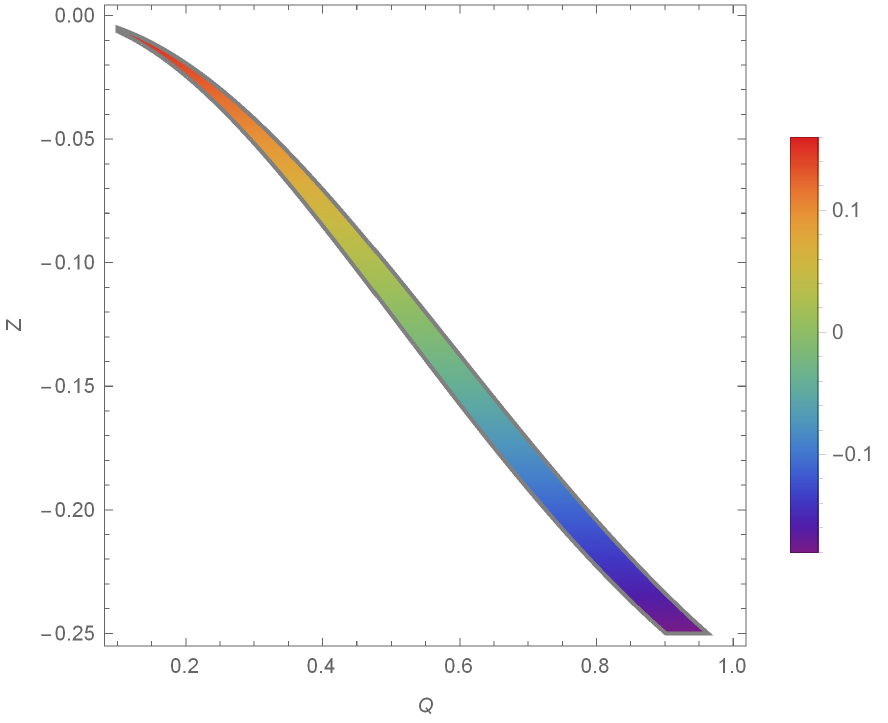}}
     \newline
\caption{\label{Fig4} The shadow diameter deviation
from a Schwarzschild BH as a function of $(Q, Z)$ for the fixed $\mu=0.4$. Note that the white
region is forbidden for $Z$ and $Q$. The mass parameter is set to be $M=1$.  }
\end{figure}

We will then consider the recent EHT observations and apply the obtained theoretical results to provide upper constraints on the BH parameters. In this regard, we can use the observational data of the supermassive BH at the center of M87$^{\star}$, which can be considered as a static and spherically symmetric charged BH for the theoretical model, supporting the assumption for our theoretical model considered here. With this in view, we obtain constraints on the upper values of the parameters $Z$ and $Q$, following the EHT observational data through the angular diameter $\theta$ of the BH shadow, the distance $D$, and the mass of the M87$^{\star}$ galaxy as the supermassive BH. The observational data provides the aforementioned observational parameters for M87$^{\star}$: $\theta_{M87^{\star}}=42 \pm 3 \mu as$, $D= 16.8 \pm 0.8 M pc$ between Earth and M87$^{\star}$ with mass $M_{M87^{\star}} = (6.5 \pm 0.7) \times 10^9 M_{\odot}$ (see, e.g., \cite{Akiyama19L1,Akiyama19L6}). We then evaluate the BH shadow diameter per unit mass using the following expression \cite{Bambi19PRD}: 
\begin{eqnarray}
    d_{sh}=\frac{D\,\theta}{M}\, .
\end{eqnarray}
Keeping $d_{sh}=2r_{sh}$ in mind, we further estimate the BH shadow diameter by imposing observational data $d^{M87^{\star}}_{sh}=(11 \pm 1.5)M$ for M87$^{\star}$. Taking the observational EHT data of M87$^{\star}$ into consideration, we demonstrate the upper values of $Z$ and $Q$ in Fig.~\ref{Fig3}. As can be seen from Fig.~\ref{Fig3}, we find constraints on the $R^{2}$ gravity parameter $Z$ and the charge $Q$ within the confidence levels of $1\sigma$ and $2\sigma$ while keeping $\mu$ fixed using the EHT observational data of M87$^{\star}$. In Fig.~\ref{Fig3}, the shaded regions represent the constraint space of $(Q,Z)$ within the confidence levels of $1\sigma$ and $2\sigma$. As inferred from Fig.~\ref{Fig3}, the positive $Z>0$ and the BH charge $Q$ support each other, i.e., the upper value of the coupling parameter $Z$ increases with an increasing upper value of the charge parameter. However, the opposite is true for the negative $Z<0$, resulting in its upper limits decreasing with an increasing upper limit of the BH charge parameter.  Fig.~\ref{Fig4} displays the shadow diameter deviation from a Schwarzschild BH ($ \delta=\frac{d_{sh}}{6\sqrt{3}}-1 $) as a function of $(Q, Z)$ while keeping the cosmological constant $\mu$ fixed.  According to reported results of M87$^{\star}$ data, the bound of the measured Schwarzschild deviation is as $-0.18<\delta <0.16$. For positive $\delta$, the black hole shadow size is greater than the Schwarzschild black hole of the same mass, and vice versa. The shaded region in Fig.~\ref{Fig4} shows the region of parameters which satisfies the mentioned constraint for $\delta$.  It can be seen from Fig.~\ref{Fig4}(a) that the deviation from a Schwarzschild BH case approaches positive values as the parameter space of $(Q,Z)$ increases, meaning that a charged AdS BH in $R^{2}$ gravity with the same mass as Schwarzschild, has a greater (smaller) shadow size than that of the Schwarzschild BH for large (small) values of the parameters $Q$ and $Z$. While charged dS BHs in $R^{2}$ gravity have a greater shadow size compared to the Schwarzschild one for small values of the parameters (see Fig.~\ref{Fig4}(b)).
\subsection{Energy emission rate}

Here, we now move to the study of the associated rate of energy emission. It is widely believed that quantum fluctuations in a BH spacetime may lead to a condition in which pairs of particles can be created and annihilated in the close environment of the BH, especially very close to the horizon. As a consequence of this process, particles with positive energy are able to escape from the BH through quantum tunneling. Therefore, the BH evaporates as Hawking-radiation over a specific period of time. For a static and spherically symmetric BH case, the absorption cross-section oscillates around a limiting constant value $\sigma_{lim}$ in the limit of very high energies. Under this process, the BH shadow results in the high-energy cross-section of absorption, which can be observed if the observer is located at a finite distance. On the other hand, it is approximately equivalent to the photon sphere area $ \sigma _{lim}\approx \pi r_{sh}^{2} $. The energy emission rate around a BH is then defined by \cite{Decanini11Cclass}
\begin{figure}[!htb]
\centering
\subfloat[$Z=0.5$ ]{
        \includegraphics[width=0.4\textwidth]{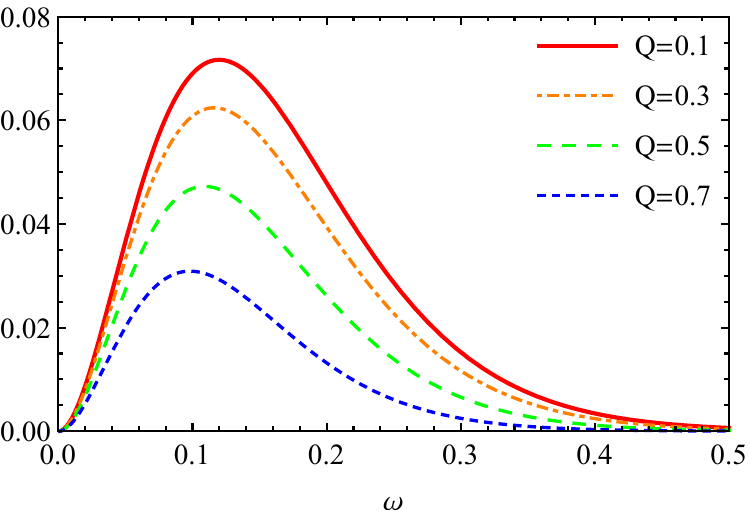}}
\subfloat[$ Q=0.2 $ ]{
     \includegraphics[width=0.4\textwidth]{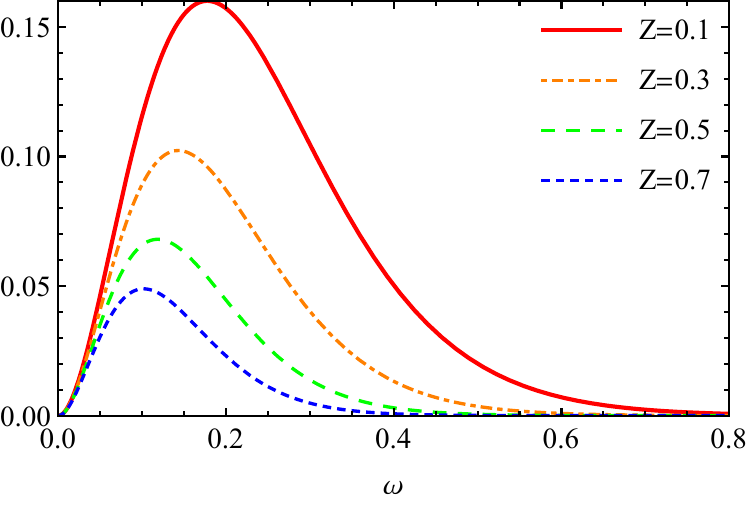}}
\begin{center}     
\subfloat[$ Q=0.2 $]{
        \includegraphics[width=0.4\textwidth]{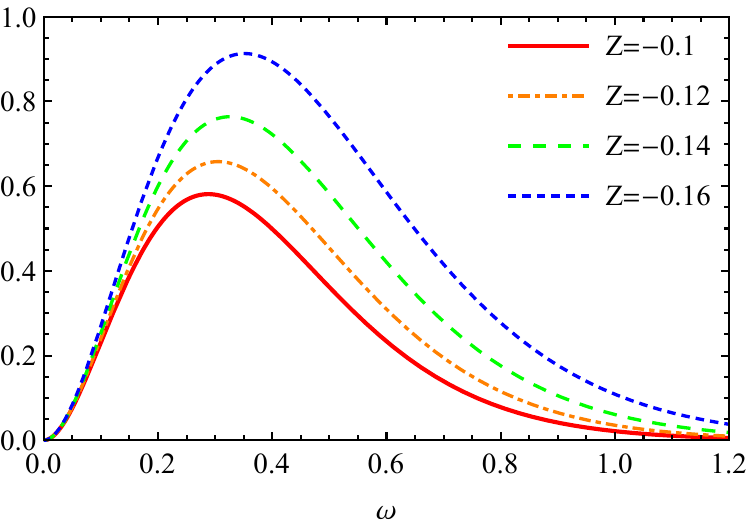}}
        \end{center}
\caption{\label{Fig5} The emission frequency profile of the energy emission rate for various combinations of the BH charge $Q$ and the coupling parameter $Z$ while keeping 
the cosmological constant $\mu=0.4$ fixed. Note that we set $M=1$. }
\end{figure} 
\begin{eqnarray}
\frac{d^{2}\mathcal{E}(\omega )}{dtd\omega }=\frac{2\pi ^{3}\omega
^{3}r_{sh}^{2}}{e^{\frac{\omega }{T}}-1}\, ,  \label{Eqemission}
\end{eqnarray}
with the emission frequency $\omega $ and the Hawking temperature $T$, which is given by
\begin{equation}
T=\frac{1}{4\pi}\left( \frac{M}{r_{e}^{2}}+\frac{2Z}{r_{e}^{3}}+\frac{\mu^{2}Q^{2}r_{e}}{3Z}\right)\, ,  \label{EqTH}
\end{equation}
in which $ r_{e} $ is the horizon radius. We now turn to examine the behavior of energy emission rate
as a function of $\omega$. To this end, we demonstrate its behavior around the charged BH in $R^2$ gravity for various combinations of parameters $Q$ and both positive and negative $Z$ in Fig.~\ref{Fig5}. The top left panel shows the impact of the BH charge $Q$ on the emission frequency profile of the energy emission rate, while the top right/bottom panel reflects the role of the coupling parameter $Z>0(<0)$ while keeping the cosmological constant $\mu$ fixed. From Fig.~\ref{Fig5}, the charge $Q$ decreases the energy emission rate, resulting in the curves shifting downward toward smaller values of the emission rate. Similarly, the curves of the energy emission rate also shift down under the influence of the coupling parameter $Z$, as shown in the top right panel of Fig.~\ref{Fig5}. Unlike $Z>0$, the energy emission rate increases with an increasing coupling parameter $Z$ for the negative case. Also, it is to be emphasized that the energy emission rate is effectively increased as a consequence of the impact of the negative $Z$, compared to the impacts of positive $Z$ and $Q$. This is a remarkable aspect of the coupling parameter in $R^2$ gravity.   However, it should be noted that the BH shadow decreases with an increasing the coupling parameter $Z<0$, similarly to what is observed in the BH shadow boundary resulting from the charge parameter $Q$. As a result, the evaporation process will be faster for a BH located in a weak electric field, meaning that neutral BHs have shorter lifetimes compared to their charged counterparts in $R^{2}$ gravity. Additionally, a charged BH in $R^{2}$ gravity evaporates very slowly/fast in AdS/dS background which reveals the fact that a black hole has a longer/shorter lifetime in AdS/dS spacetime.

\section{\label{Sec:rot_bh} A rotating charged BH in $R^{2}$ gravity}

\begin{figure}[!htb]
\centering
\subfloat[]{
        \includegraphics[width=0.44\textwidth]{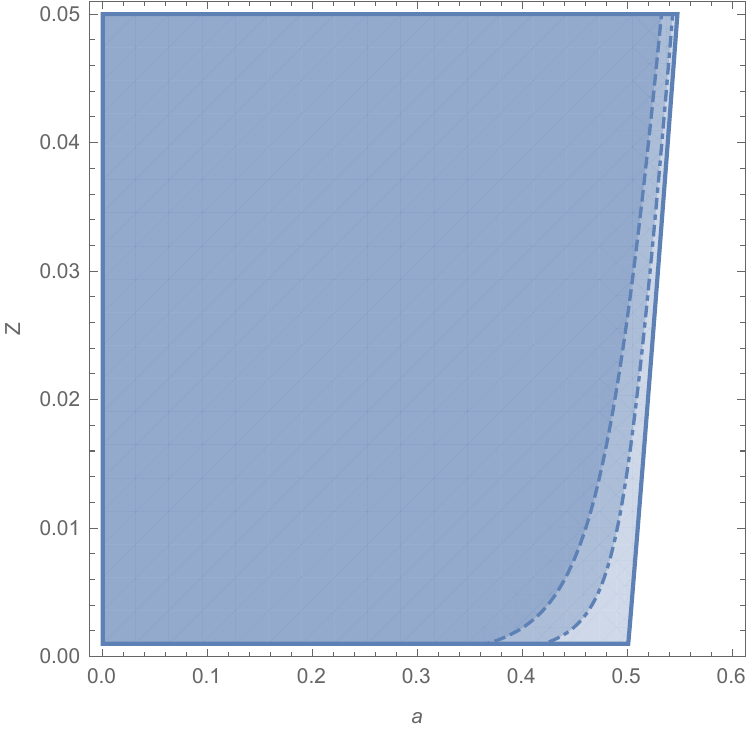}}
\subfloat[]{
     \includegraphics[width=0.45\textwidth]{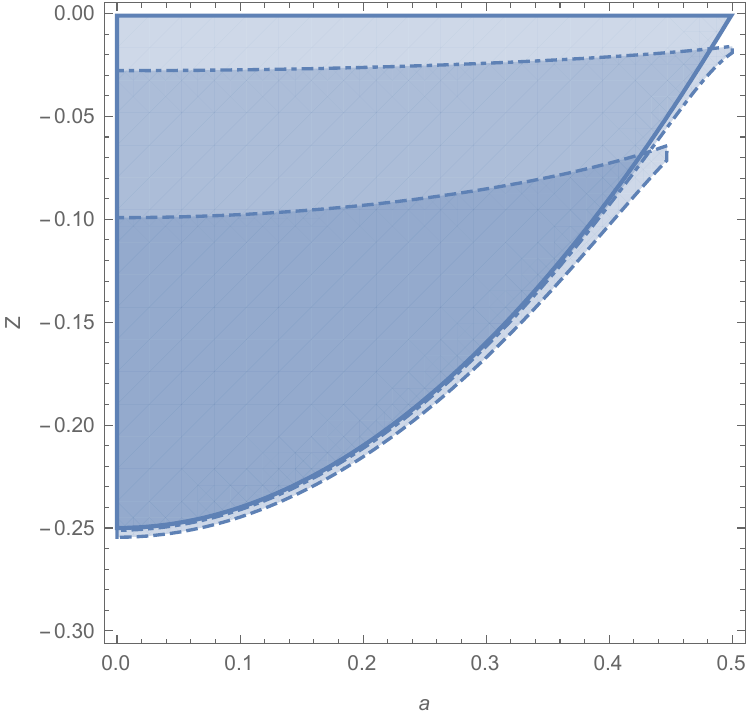}}
\caption{\label{Fig6} The admissible parameter space plot (shaded areas) between the spin parameter $a$ and the coupling parameter $Z$ for various combinations of the BH charge parameter $Q=0.01,\, 0.4,\, 0.8$ corresponding to solid, dash-dotted, and dashed curves, respectively. Note that the white area refers to the region, where no black hole solution exists. 
}
\end{figure}
\begin{figure}[!htb]
\centering
\subfloat[]{
        \includegraphics[width=0.31\textwidth]{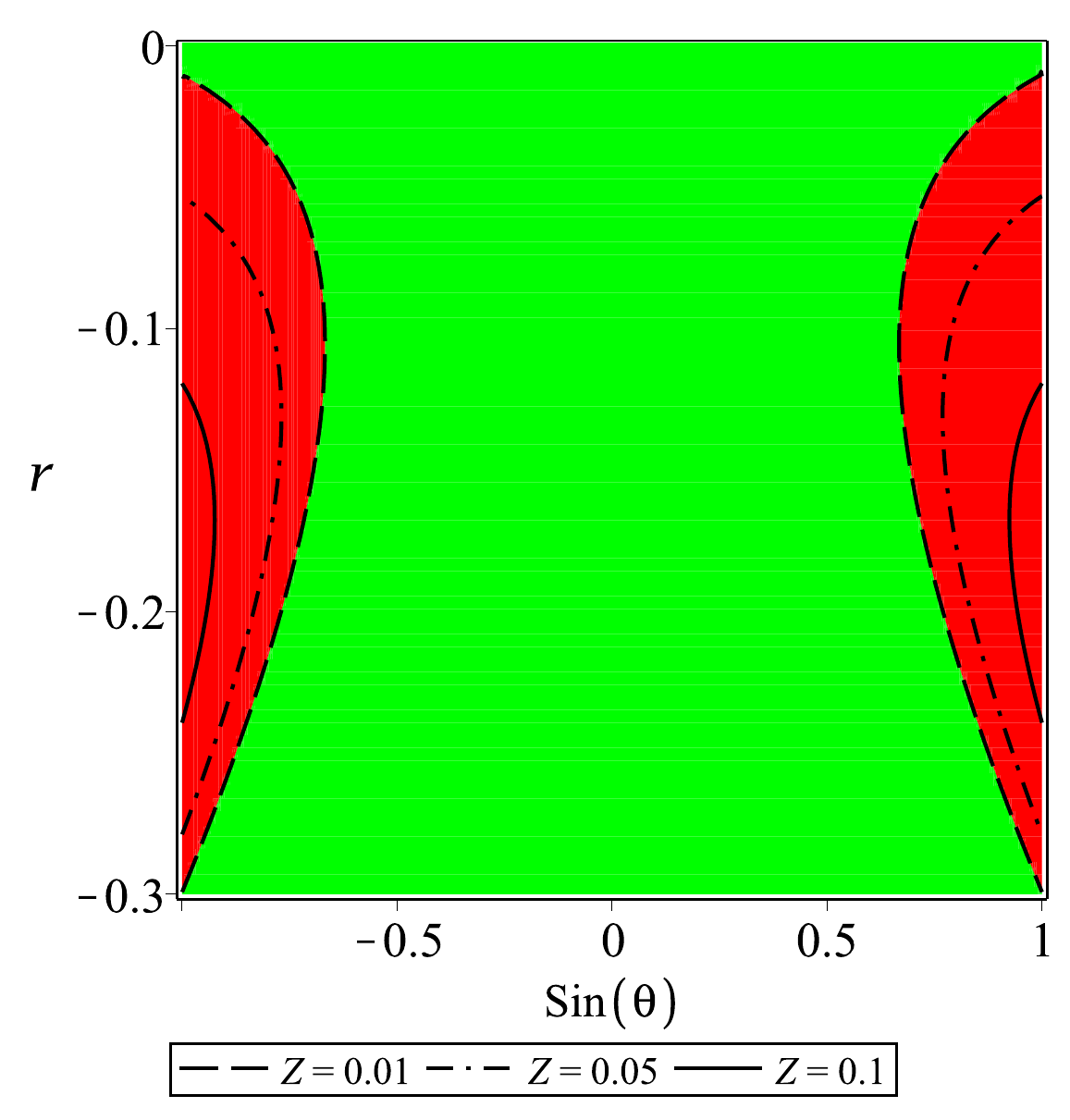}}
\subfloat[]{
     \includegraphics[width=0.31\textwidth]{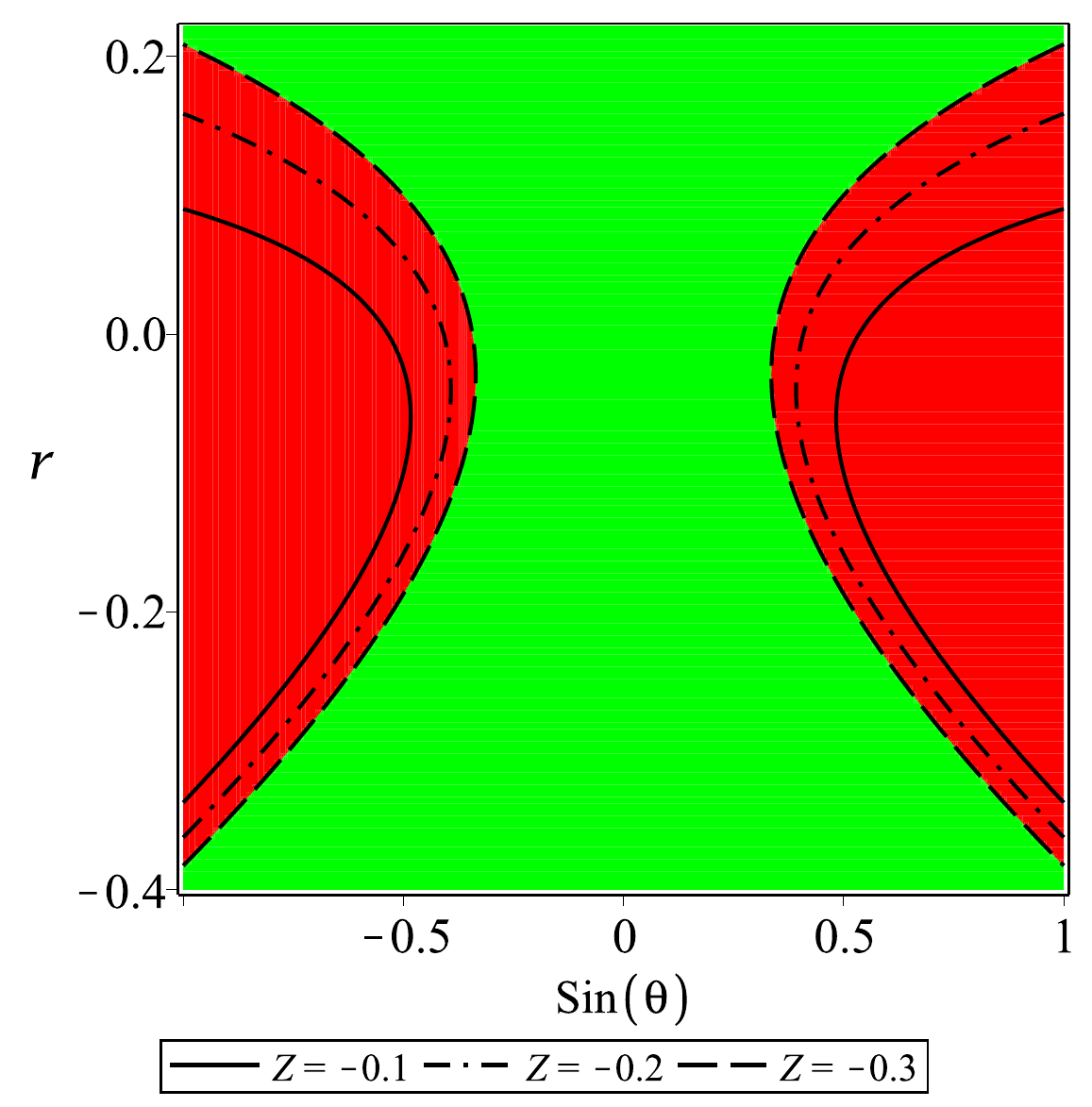}}
\subfloat[]{
     \includegraphics[width=0.33\textwidth]{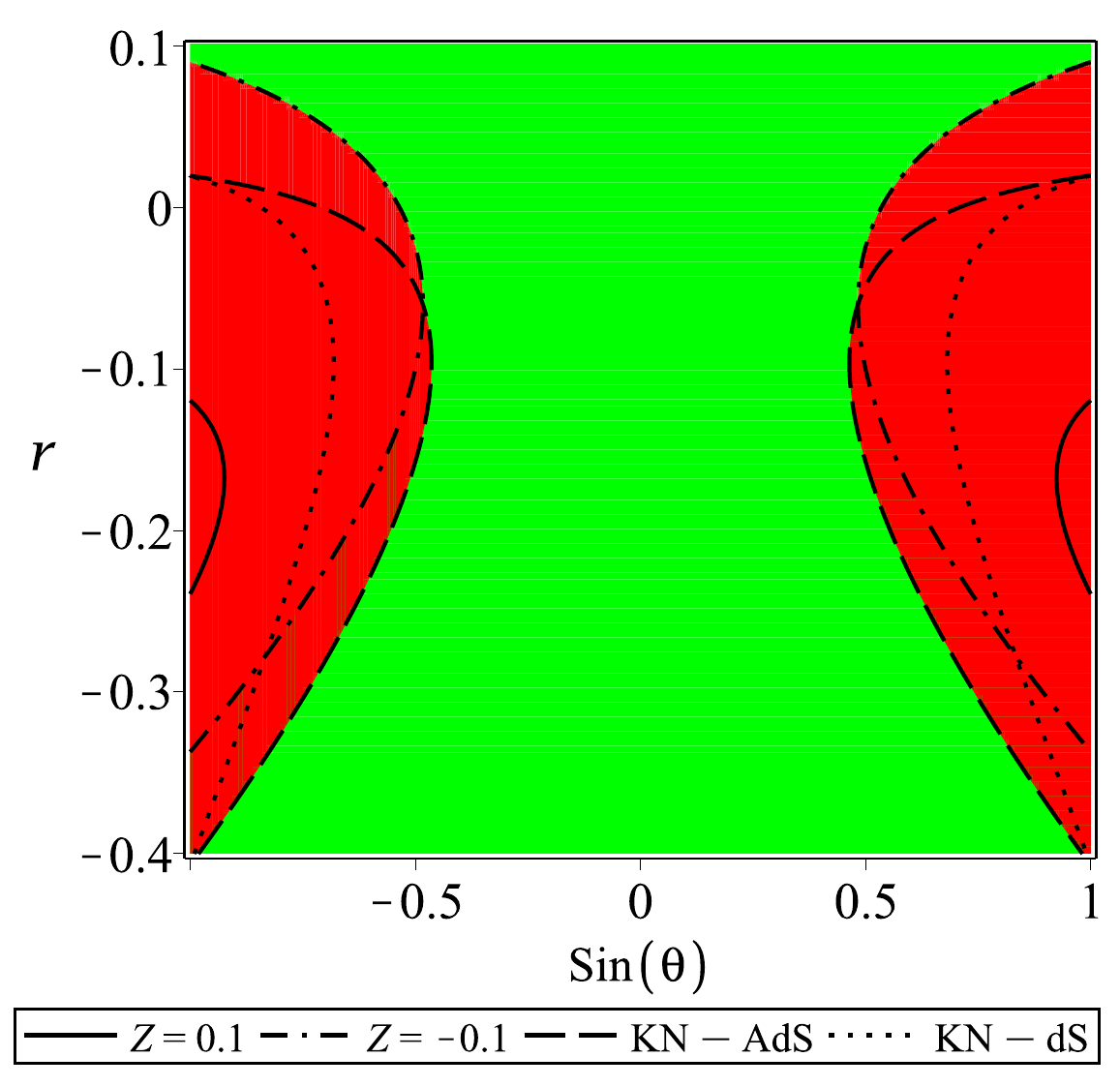}}     
     \newline
\caption{The Causality violations/preservation regions of the black hole. Here we have fixed
$M = 1$, $ Q=0.2 $ and $ \mu=0.4 $.}
\label{Fig7}
\end{figure}

In this section, we generalize the static and spherically symmetric charged BH solution in
$R^{2}$ gravity to the Kerr-like rotating BH solution by adapting the Newman-Janis algorithm (NJA) \cite{Azreg14CPRD,Azreg14EPJC}. It is valuable to note that the final result in Eddington-Finkelstein coordinates (EFC) may not be transformed into the Boyer-Lindquist coordinates (BLC) due to the complexification process when applying this method to other spacetime metrics. This is a remarkable aspect of the NJA algorithm. Also, it is worth noting that it is possible for the rotating spacetime obtained by the NJA within the EFC to be transformed into the BLC without taking the complexification process into consideration. To apply NJA without complexification, we consider a general form of a static and spherically symmetric spacetime as follows
\begin{equation}
ds^2=-A(r)dt^2+ \frac{dr^2}{B(r)} + C(r)d\Omega^2\, .
\label{general_metric}
\end{equation}
with $d\Omega^2 = d\theta^2 + \sin^2\theta d\phi^2$. One can then be able to transform the given metric Eq.~(\ref{general_metric}) from the BLC $(t,r,\theta,\phi)$
to EFC $(u,r,\theta,\phi)$ by imposing the following coordinate transformation:
\begin{equation}
du=dt-\frac{dr}{\sqrt{A(r) B(r)}}\, .
\end{equation}
The spacetime metric (i.e., Eq.~(\ref{general_metric})) does therefore take the form as
\begin{eqnarray}
ds^2=-A(r)du^2-2\sqrt{\frac{A(r)}{B(r)}}dudr + C(r)d\Omega^2\, ,
\label{general_metric_null}
\end{eqnarray}
and its null tetrads are written as 
\begin{eqnarray}
&& l^\mu=\delta^\mu_1\,, \label{l_mu} \\
&& n^\mu=\sqrt{\frac{B(r)}{A(r)}}\delta^\mu_0-\frac{1}{2}B(r)\delta^\mu_1\, , \label{n_mu} \\
&& m^\mu=\frac{1}{\sqrt{2C(r)}} \left(\delta^\mu_2+\frac{i}{\sin\theta}\delta^\mu_3 \right)\, . \label{m_mu}
\end{eqnarray}

The next step is the complexification, which
generalizes the functions $A(r)$, $B(r)$, and $C(r)$ as real functions of the radial coordinate $r$ and its complex conjugate $\bar{r}$. The trick is to skip this step and move on to the next, which leads to the requirement of complex coordinate transformations
\begin{eqnarray}
&& r^{\prime}=r+ia\cos \theta\, ,
\label{r new}\\
&& u^{\prime}=u-ia\cos \theta\, .
\label{u new}
\end{eqnarray}
As a consequence of these transformations, the metric functions then take on new forms. Substituting the new metric functions into the line element, as given by Eq.~(\ref{general_metric_null}), results in a complicated metric form. This can be avoided by transforming the null coordinates $(u,r,\theta,\Phi)$ into the Boyer-Lindquist coordinates (BLC) $(t,r,\theta,\phi)$ using the following transformation: 
\begin{eqnarray}
&& du=dt-\lambda(r)dr\, ,
\label{du new}\\
&& d\Phi=d\phi-\chi(r)dr\, ,
\label{dphi new}
\end{eqnarray}
where functions, $\chi(r)$ and $\lambda (r)$, are defined as   
\begin{eqnarray}
&& \chi(r)=\frac{a}{ CB +a^2}\, ,
\label{chi}\\
&& \lambda(r)=\frac{K+a^2}{CB+a^2}\, ,
\label{lambda}
\end{eqnarray}
with $K=C\sqrt{\frac{B}{A}}$. In the Boyer-Lindquist coordinates ($t,r,\theta,\phi$), the metric can then be rewritten as  
\begin{equation}\label{eq:rotating Kerr}
  \begin{split}
  \mathrm{d}s^{2}=&-\frac{\left(BC+a^{2}\cos^{2}{\theta}\right)\Sigma}{\left(C+a^{2}\cos^{2}{\theta}\right)^{2}}\mathrm{d}t^{2}+\frac{\Sigma}{BC+a^{2}}\mathrm{d}r^{2}-2a\sin^{2}\theta\left[\frac{C-BC}{\left(C+a^{2}\cos^{2}{\theta}\right)^{2}}\right]\Sigma\mathrm{d}\phi\mathrm{d}t\\
  &+\Sigma\mathrm{d}\theta^{2}+\Sigma\sin^{2}\theta\left[1+a^{2}\sin^{2}\theta\frac{2C-BC+a^{2}\cos^{2}\theta}{\left(C+a^{2}\cos^{2}{\theta}\right)^{2}}\right]\mathrm{d}\phi^{2}\, .
  \end{split}
  \end{equation}
Taken altogether, we can write the form of the spacetime metric of rotating BHs in $R^{2}$ gravity as:
\begin{eqnarray} \label{general_rotating_metric}
  ds^2 &=& -\left(\frac{\Delta(r)-a^{2}\sin^{2}\theta}{\Sigma(r,\theta)} \right)dt^2 + \frac{\Sigma(r,\theta)}{\Delta(r)} dr^2 + \Sigma(r,\theta) d\theta^2 +\frac{\sin^2\theta\left( \left( a^{2}+r^{2}\right)^{2}-\Delta(r)a^{2}\sin^2\theta \right) }{\Sigma(r,\theta)} d\phi^2  \\ \nonumber
 & -&\,  \frac{2a\sin^2\theta \left(a^{2}+r^{2}-\Delta(r) \right) }{\Sigma(r,\theta)} dt d\phi\, . 
\end{eqnarray}
where we have used  
\begin{eqnarray}
   && \Delta(r)=r^{2}-Mr-Z+a^{2}+\frac{\mu^{2}Q^{2}r^{4}}{6Z}\, , \\
   && \Sigma(r,\theta) = r^{2} + a^2\cos^2\theta\, .
\end{eqnarray}
The rotating solution, as given by Eq.~(\ref{general_rotating_metric}), reduces to the static and spherically symmetric case in the limit of $a \to0$. To determine the allowed regions of the parameters for a valid physical BH solution, we demonstrate the admissible parameter space plot ($a-Z$) for various combinations of the BH charge parameter $Q$ in Fig.~\ref{Fig6}. The admissible parameter space region shows that its allowed region shrinks as a consequence of an increase in the BH charge $Q$ for both positive and negative cases of the coupling parameter $Z$; see Fig.~\ref{Fig6}. Interestingly, the negative case of the coupling parameter $Z<0$ in $R^2$ gravity leads to larger white regions where BHs cannot exist, compared to its opposite case, as shown in the right panel of Fig.~\ref{Fig6}.   

One concern with studying Kerr-like BHs is the existence of closed timelike curves (CTCs). CTCs are trajectories in spacetime that effectively travel backwards in time. In fact, the existence of CTCs violets the causality relations, while the formulation of the fundamental laws of physics is based on the preservation of causality. Causality violations and CTCs are possible if $ g_{\phi\phi} >0$. To further study this issue, we plot Fig.~\ref{Fig7}, which shows a plot of $r$ versus $\sin\theta$. The red-colored region represents where $g_{\phi\phi} >0$, indicating that 
causality violations and CTCs occur in this region. The green-colored region is where causality relations are preserved. From Fig.~\ref{Fig7} (a), it can be seen that the KN-AdS like BH in $ R^{2} $ gravity has CTCs for $r<0$, indicating that there are no causality violations for this BH. While the KN-dS like BH has CTCs for both signs of $r$, which shows that the causality relations can be preserved for particular choices of the parameters (see e.g., Fig. \ref{Fig7} (b)). A comparison between the KN-AdS/dS BH and KN BHs in $R^{2}$ gravity is depicted in Fig.~\ref{Fig7} (c). From this figure, it is clear that the smallest/biggest region of CTCs is related to KN-dS in GR/KN-dS in $R^{2}$ gravity.

\subsection{Null geodesics and BH shadow}
In this subsection, we consider null geodesics of rotating BH spacetime in $R^2$ gravity, as shown in Eq.~(\ref{general_rotating_metric}). For that we find the contour of the BH shadow related to the rotating spacetime by separating the null geodesic equations using
the Hamilton-Jacobi formalism, which is given by \cite{Hendi23}
\begin{equation}
H=-\frac{\partial S}{\partial \sigma}=\frac{1}{2}g^{\mu\nu}\frac{\partial S}{\partial x^{\mu}}\frac{\partial S}{%
\partial x^{\nu}}=0\, ,
\label{HJ1}
\end{equation}
where $ H $ and $S$ are the canonical Hamiltonian and the Jacobi action, respectively. $\sigma$ refers to the affine parameter along the geodesics. Using known constants of the motion, the Jacobi action can be separated as
\begin{equation}
S=-\frac{1}{2}m^2\sigma+Et-L\phi+S_r(r)+S_{\theta}(\theta),
\label{HJ2}
\end{equation}
where $m$ is the test particle mass, and $S_r(r)$ and $S_\theta(\theta)$ are the function of $r$ and $\theta$. Moreover, $E=p_{t}$ and $L= p_{\phi}$ denote the conserved energy and angular
momentum, respectively. Combining
Eq.~(\ref{HJ1}) and Eq.~(\ref{HJ2}) and applying the variable
separable method, the null geodesic equations for the photon ($m=0$) are obtained as 
\begin{align}
\label{HJ3}
&\Sigma\frac{dt}{d\sigma}=\frac{r^2+a^2}{\Delta_{r}}\left( E(r^2+a^2)-aL\right) -a^{2}E\sin^2\theta-aL),\\
&\Sigma\frac{dr}{d\sigma}=\sqrt{\mathcal{R}(r)},\\
&\Sigma\frac{d\theta}{d\sigma}=\sqrt{\Theta(\theta)},\\
\label{HJ4}
&\Sigma\frac{d\phi}{d\sigma}=\frac{a}{\Delta_{r}}[E(r^2+a^2)-aL]-\left(aE-\frac{L}{\sin^2\theta}\right),
\end{align}
in which $\mathcal{R}(r)$ and $\Theta(\theta)$ are given by
\begin{align}
\label{HJ5}
&\mathcal{R}(r)=[E(r^2+a^2)-aL]^2-\Delta_{r}[(aE-L)^2+\mathcal{K}],\\
&\Theta(\theta)=\mathcal{K}+a^2E^{2}\cos^2\theta-L^2\cot^2\theta,
\end{align}
where $\mathcal{K}$ is a constant of separation called Carter constant. For $\mathcal{K}=0$,
$\theta$-motion is suppressed, and all photon orbits are
restricted only to the equatorial plane ( i.e., $\theta=\pi/2 $), yielding unstable
circular orbits at the equatorial plane. In order to obtain the boundary of the shadow, we can express the radial geodesic
equation in terms of effective potential $V_{\text{eff}}$ of photon’s radial motion as
\begin{equation*}
\Sigma^2\left(\frac{dr}{d\sigma}\right)^2+V_{\text{eff}}=0.
\end{equation*}
\begin{figure}[!htb]
\centering
\subfloat[$ a=0.5 $, $ Z=0.2 $ and $\theta=\pi /2$]{
        \includegraphics[width=0.32\textwidth]{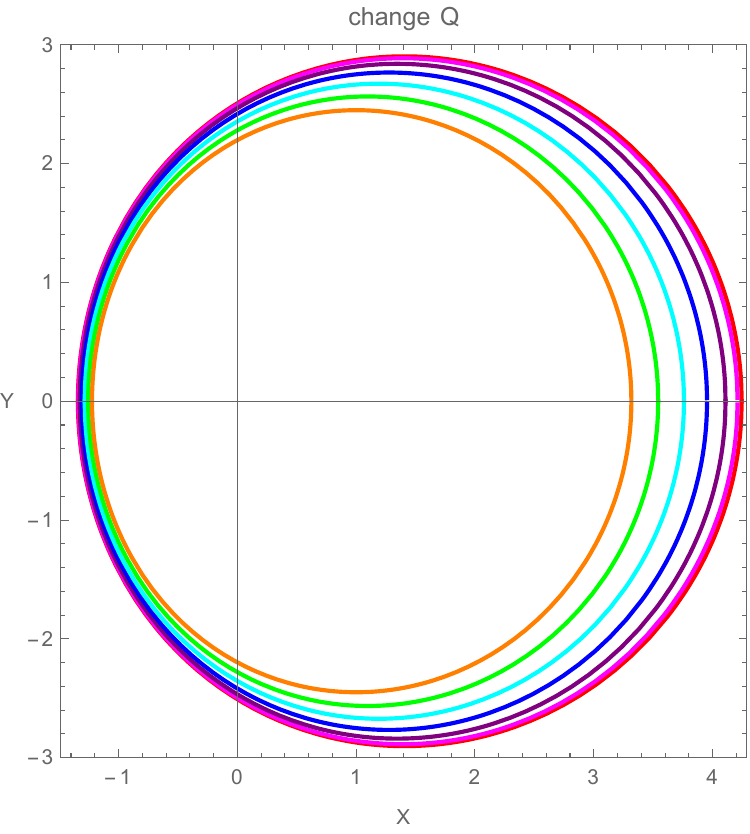}}
\subfloat[$ a=0.5 $, $ Q=0.2 $ and $\theta=\pi /2$]{
     \includegraphics[width=0.3185\textwidth]{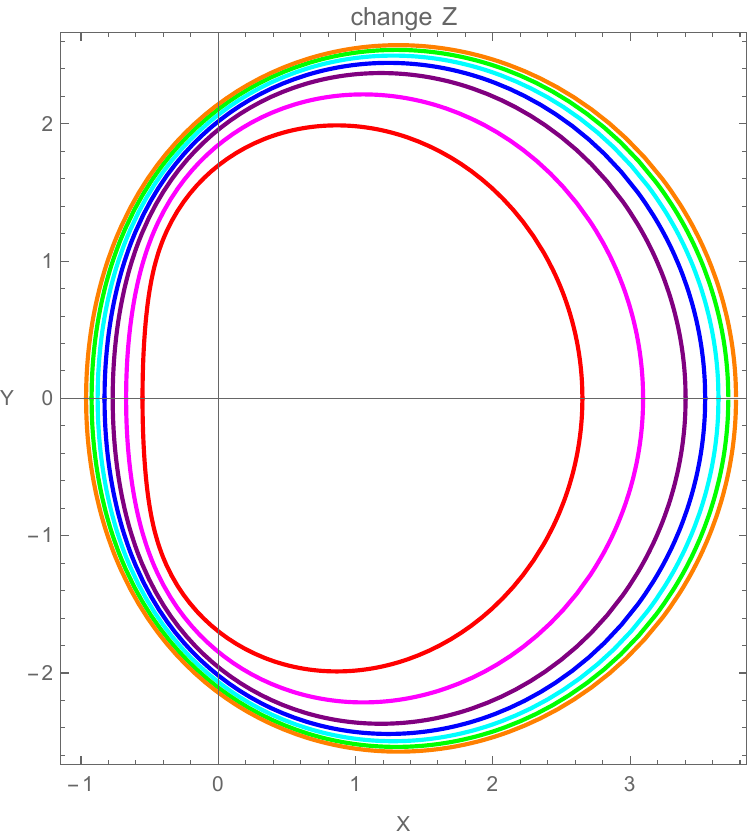}}
     \newline
\subfloat[$ Q=Z=0.2 $ and $\theta=\pi /2$]{
        \includegraphics[width=0.345\textwidth]{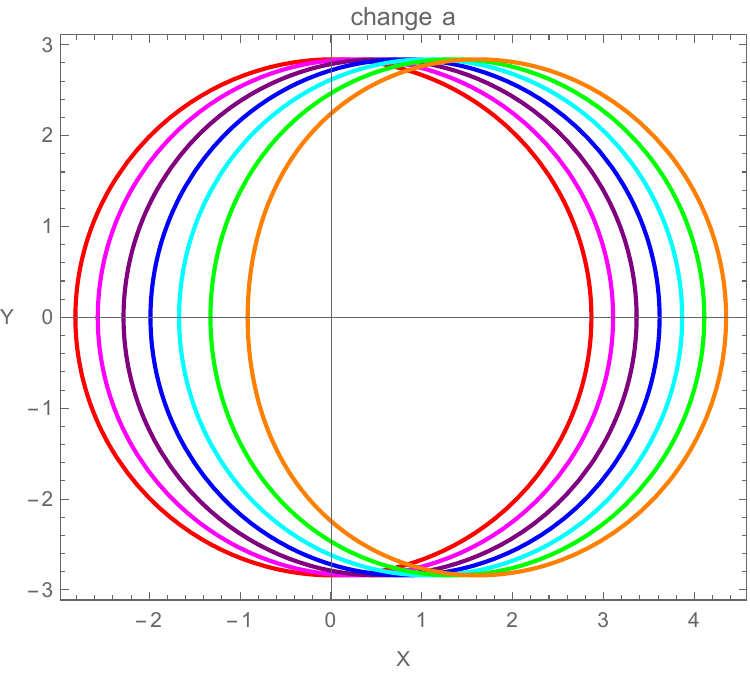}}
\subfloat[$ a=0.5 $ and $ Q=Z=0.2 $]{
        \includegraphics[width=0.34\textwidth]{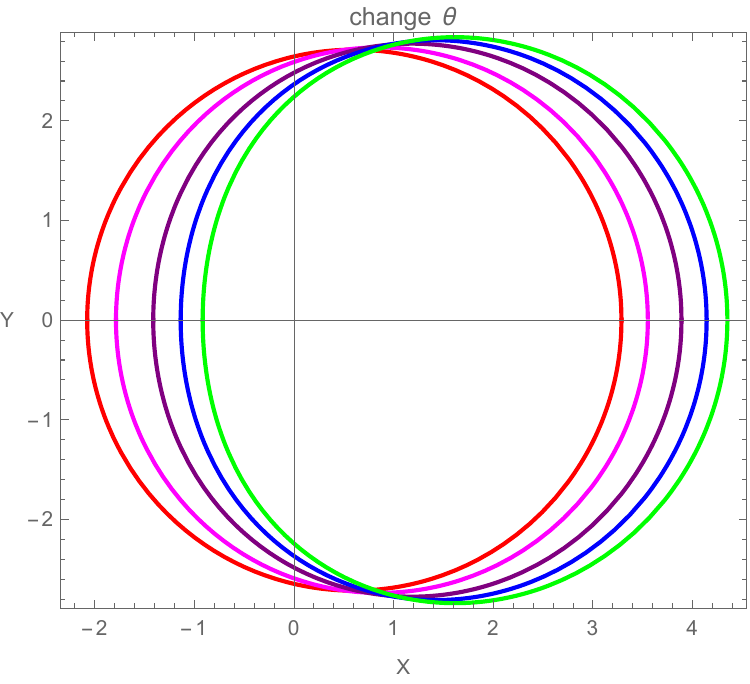}}        
\newline
\caption{\label{Fig8} The dependence of the boundary of the BH shadow on the parameters $Q$ and $Z$ and $a$ while keeping
$\mu=0.4$ fixed. {Left panel in the top row:} The boundary of the BH shadow is plotted for various values of the parameter $Q$ ranging from 0.1 to 0.6 with equal intervals, corresponding to the curves from outermost (red) to the innermost (orange). {Right panel in the top row:} The boundary of the BH shadow is plotted for various values of the parameter $Z$ ranging from 0.1 to 0.7 with equal intervals, corresponding to the curves from innermost (red) to the outermost (orange). {Left panel in the bottom row:} The boundary of the BH shadow is plotted for various values of the parameter $a$ ranging from 0.1 to 0.7 with equal intervals, corresponding to the curves from innermost (red) to the outermost (orange). {Right panel in the bottom row:} The boundary of the BH shadow is plotted for various values of the inclination angle $\theta$ ranging from $\pi/9$ to $\pi/2$ with equal intervals, corresponding to the curves from innermost (red) to the outermost (green). 
}
\end{figure}
Introducing two impact parameters $\xi$ and $\eta$
\begin{equation}
\xi=L/E,  \;\; \;\; \;\;   \eta=\mathcal{K}/E^2\, ,
\end{equation}
we rewrite $V_{\text{eff}}$ in terms of $\xi$ and $\eta$ as
\begin{equation}\label{veff}
V_{\text{eff}}=\Delta_{r}((a-\xi)^2+\eta)-(r^2+a^2-a\;\xi)^2\, ,
\end{equation}
where we have used $V_{\text{eff}}/E^2\to V_{\text{eff}}$. It is well-established fact that the shadow boundary is mainly determined by the circular photon orbit, satisfying the following conditions
\begin{equation}\label{cond}
V_{\text{eff}}(r_{ph})=0 \quad\mbox{and}\quad\frac{dV_{\text{eff}}(r_{ph})}{dr}=0\, ,
\end{equation}
whereas instability of orbits obeys the condition
\begin{equation}
\frac{d^{2}V_{\text{eff}}(r_{ph})}{dr^{2}}\leq 0\, ,
\end{equation}
Solving Eq.~(\ref{cond}) simultaneously, one can obtain the critical values of impact
parameters for unstable orbits of photons, and they are  
\begin{eqnarray}
\xi_{c}&=& \frac{(r_{ph}^{2}+a^{2})\Delta'(r_{ph})-4r_{ph}\Delta (r_{ph})}{a\Delta'(r_{ph})}\, ,\\
\eta_{c}&=& \frac{r_{ph}^{2}\left[ 16a^{2}\Delta (r_{ph})+8r_{ph}\Delta (r_{ph})\Delta'(r_{ph})-16\Delta (r_{ph})^{2}-r_{ph}^{2}\Delta'(r_{ph})^{2}\right] }{a^{2}\Delta'(r_{ph})^{2}}\, .
\end{eqnarray}
Taking Eq.~(\ref{celestial:1a}) together with the null
geodesic Eqs.~(\ref{HJ3}-\ref{HJ4}) into consideration, it is then straightforward to obtain the relations between celestial coordinates and impact parameters as follows \cite{Vazquez04Nuovo}
\begin{eqnarray}
\label{celestial:1a}
x&=&\lim_{r_0\rightarrow\infty}\left(-r_0^2\sin\theta_0\frac{d\varphi}{dr}\Big|_{(r_0,\theta_0)}\right)\, ,\\ \nonumber
y&=&\lim_{r_0\rightarrow\infty}\left(r_0^2\frac{d\theta}{dr}\Big|_{(r_0,\theta_0)}\right)\, ,
\end{eqnarray}
giving 
\begin{equation}
x=\frac{a\sin\theta_0 -\xi\csc\theta_0}{\sqrt{1-\frac{\mu^{2}Q^{2}}{6Z}\left[(a-\xi)^{2}+\eta \right] }}\, ,\qquad
y=\pm \sqrt{\frac{\eta+a^2\cos^2\theta_0-\xi^2\cot^2\theta_0}{\sqrt{1-\frac{\mu^{2}Q^{2}}{6Z}\left[(a-\xi)^{2}+\eta \right] }}}\, .
\label{xy}
\end{equation}
We now analyze the BH shadow boundary resulting from the BH charge $Q$, coupling parameter $Z$, and the spin parameter $a$  within the modified theory of $R^2$ gravity. In Fig.~\ref{Fig8}, we demonstrate the boundary of the BH shadow around the charged rotating BH in $R^2$ gravity for various combinations of BH parameters in $R^2$ gravity. In the top left panel, we show the impact of the charge parameter $Q$ on the BH shadow boundary, while in the top right panel, we show the impact of the coupling parameter $Z$ for the fixed cosmological constant $\mu$. Similarly, we show the impact of the spin parameter $a$ on the BH shadow boundary in the bottom left panel and the impact of the inclination angle $\theta$ in the bottom right panel. From Fig.~\ref{Fig8}, the charge $Q$ decreases the size of the shadow boundary, causing its shape to shrink with an increasing $Q$ ranging from 0.1 to 0.6. On the other hand, the BH shadow boundary increases under the influence of the coupling parameter $Z$ in $R^2$ gravity, resulting in the shadow sphere possibly becoming larger, as shown in the top right panel of Fig.~\ref{Fig8}. For varying values of the spin parameter $a$, an expected effect occurs where the symmetric form of the BH shadow breaks down as the horizontal diameter becomes smaller compared to the vertical diameter with increasing spin parameter $a$; as shown in the left panel in the bottom row. Interestingly, a similar effect is observed for varying values of the inclination angle $\theta$. The symmetric form of the BH shadow disappears as $\theta$ approaches $\pi/2$, resulting in an asymmetric BH shadow, as shown in the bottom right panel. This feature does however disappear when $\theta\to 0$ for a fixed $a$, as well as when $a\to 0$ for a fixed $\theta$.

\subsection{Energy Emission rate}

\begin{figure}[!htb]
\centering
\subfloat[$Q=Z=0.2$ ]{
        \includegraphics[width=0.31\textwidth]{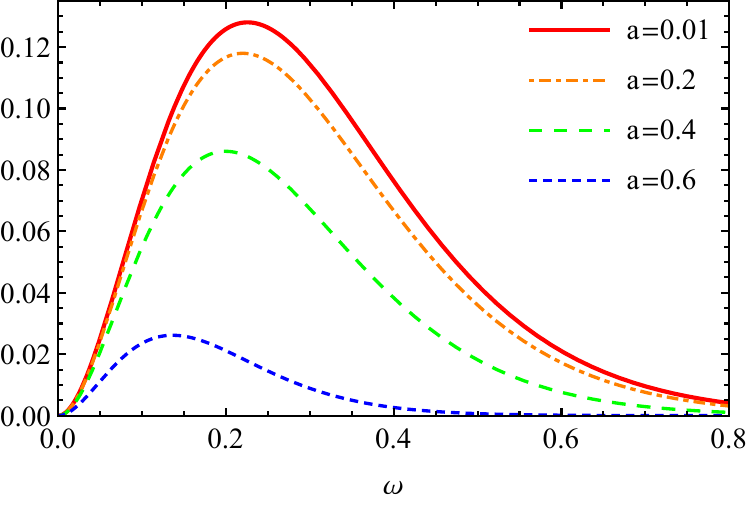}}
\subfloat[$ a=0.5 $ and $ Z=0.2 $ ]{
     \includegraphics[width=0.31\textwidth]{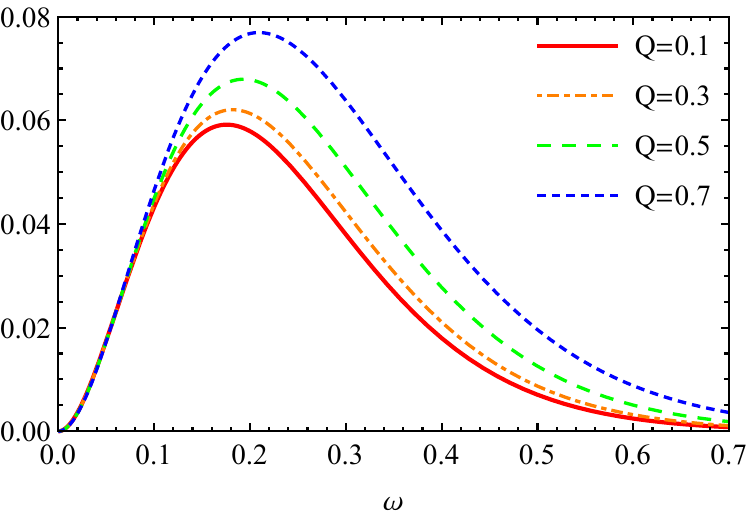}}
\newline
\subfloat[$ a=0.5 $ and $ Q=0.2 $ ]{
        \includegraphics[width=0.31\textwidth]{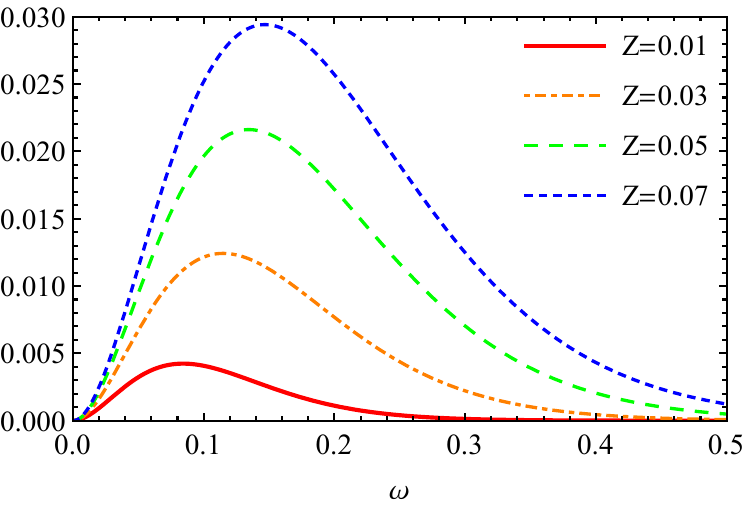}}
        \subfloat[$ a=0.5 $ and $ Q=0.2 $ ]{
        \includegraphics[width=0.31\textwidth]{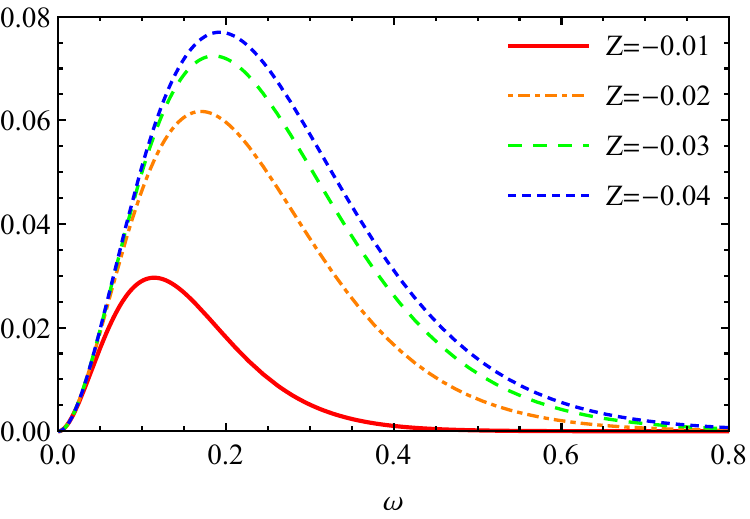}}
\newline
\caption{The emission frequency profile of the energy emission rate for the rotating charged BH in $R^2$ gravity for various combinations of the spin parameter $a$, the charge $Q$, and the coupling parameter $Z$ while keeping 
the cosmological constant $\mu=0.4$ fixed. Note that we set $M=1$. 
}
\label{Fig9}
\end{figure}
Here, we examine the behavior of energy emission rate
as a function of $\omega$ around the rotating charged BH in $R^2$ gravity. Following Eq.~(\ref{Eqemission}) and obtaining the Hawking temperature of the rotating charged BH in $R^{2}$ gravity at the horizon $r_{e}$
\begin{equation}
T=\frac{6Zr_{e}-3MZ+2\mu^{2}Q^{2}r_{e}^{3}}{12\pi Z(a^{2}+r_{e}^{2})}\, ,
\end{equation}

In this analysis, we examine the energy emission rate around the rotating charged BH. In Fig.~\ref{Fig9}, we illustrate the behavior of the energy emission rate for varying BH parameters, such as the spin parameter $a$, the charge $Q$ and both positive and negative cases of the coupling parameter $Z$ in $R^2$ gravity. Fig~\ref{Fig9}(a) depicts the influence of the rotation parameter on the energy emission rate. It is evident from this figure that the rotation parameter decreases the energy emission rate, revealing the fact  that the evaporation process would be slower for fasr-rotating BHs.  The effect of the electric charge on the emission rate is depicted in Fig~\ref{Fig9}(b). As we see, the energy emission rate increases with an increase in the electric charge, meaning that KN-AdS black holes in $R^2$ gravity evaporates very fast in the presence of a more powerful electric field. Regarding the effect of the coupling parameter $Z$,  bottom panels of Fig~\ref{Fig9} show that increasing this parameter leads to increasing energy emission in both AdS and dS spacetime. In other words, the black hole has a shorter lifetime when the effect of $R^2$ gravity gets stronger. In short, it can be said that fast-rotating BHs or black holes located in a weak electric field have longer lifetime.

\subsection{Constraints from the EHT observations of M87*}
In this subsection, we consider M87* as a KN-dS/AdS black holes in $R^{2}$ gravity and constrain the BH parameters using EHT observation. To find the the allowed regions of parameters, we need to define observables associated with black hole shadow. The first important observable is the angular size of shadow, which as was already mentioned in Subsec. \ref{IIIB}, can be derived from \cite{Bambi19PRD}
\begin{equation}
d_{M87^{*}}\equiv \frac{D\theta}{M}\approx 11.0 \pm 1.5.
\label{EqdM87b}
\end{equation}

Other Observables used to distinguish the shadow of a black hole in modified gravity to that of the Kerr scenario in GR are deviation from circularity $ \Delta C $ and the axis ratio $ D_{x} $.  In contrast to non-rotating BHs, the shadow structure for Kerr-like solution is a deformed circle and hence it is significant to define the circularity deviation $ \Delta C $ which measures the deviation from a perfect circle. The deviation from circularity
is defined as \cite{Bambi19PRD}
\begin{equation}\label{eq-DeltaC}
{\Delta C=\frac{1}{\bar{\mathbb{R}}}\sqrt{\frac{1}{2\pi}\int^{2\pi}_{0}(\mathbb{R}(\phi)-\bar{\mathbb{R}})^2 d\phi}},
\end{equation}
in which the average radius of the shadow $ \bar{\mathbb{R}} $ has the form 
\begin{equation}
\bar{\mathbb{R}}^2=\frac{1}{2\pi}\int^{2\pi}_{0} \mathbb{R}(\phi)^2d\phi.
\end{equation}
where
\begin{equation}
\begin{split}
\mathbb{R}(\phi)=\sqrt{(x-x_c)^2+(y-y_c)^2}~~~~,
\phi=\tan^{-1}\left(\frac{y-y_c}{x-x_c}\right),
\end{split}
\end{equation}

Note that the geometric center of the black hole shadow is determined by the edges of the shaped boundary via $({x_c=\frac{x_r+x_l}{2}}, y_c=0)$. Here the subscripts $l$ and $r$,  stand for the left and right ends of the shadow silhouette. The axis ratio is given as \cite{Jafarzade24PDU}
\begin{equation}\label{eq-Dx}
D_x=\frac{y_t-y_b}{x_r-x_l}.
\end{equation}
where $ y_t $ and $ y_b $ denote the top and bottom of the reference circle. According to reported results, the EHT observations constrain the deviation from circularity as $ \Delta C \lesssim 0.1 $ and the axis ratio as $1 < D_{x} \lesssim 4/3$. 

Another important observable is deviation $(\delta)$ which measures the difference between the model shadow diameter ($ d_{metric} $) and the Schwarzschild shadow diameter defined as \cite{Jafarzade24PDU}
\begin{equation}\label{eq-Dx}
\delta=\frac{d_{metric}}{6\sqrt{3}}-1,
\end{equation}
in which $ d_{metric}=2\mathcal{R}_{s} $ with $ \mathcal{R}_{s}=\sqrt{A/\pi} $ and $ A $ is given by
\begin{equation}
A=2\int y(r_{p})dx(r_{p})=2\int_{r^{-}_{p}}^{r^{+}_{p}}\left(y(r_{p})\frac{dx(r_{p}) }{dr_{p}}\right)dr_{p}.
\label{AD}
\end{equation}

It is clear that $ \delta $ can be negative (positive) if the black hole shadow size is smaller (greater) than the Schwarzschild black hole of the same mass. Based on the EHT observations, the bound of the measured Schwarzschild deviation is as $ \delta=-0.01 \pm 0.17 $.

\begin{figure}[!htb]
\centering
\subfloat[]{
        \includegraphics[width=0.31\textwidth]{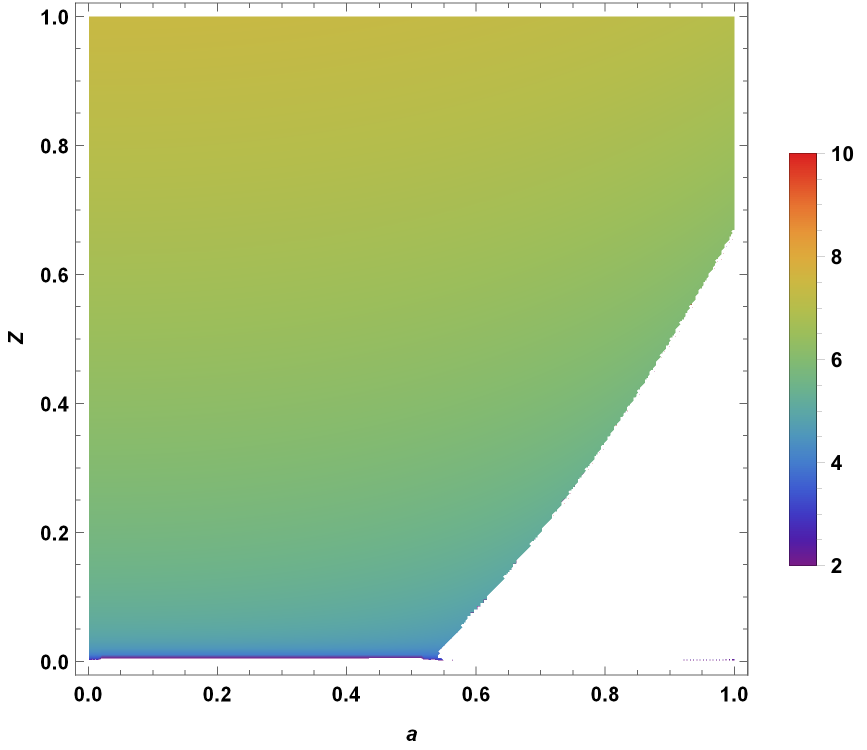}}
\subfloat[]{
     \includegraphics[width=0.32\textwidth]{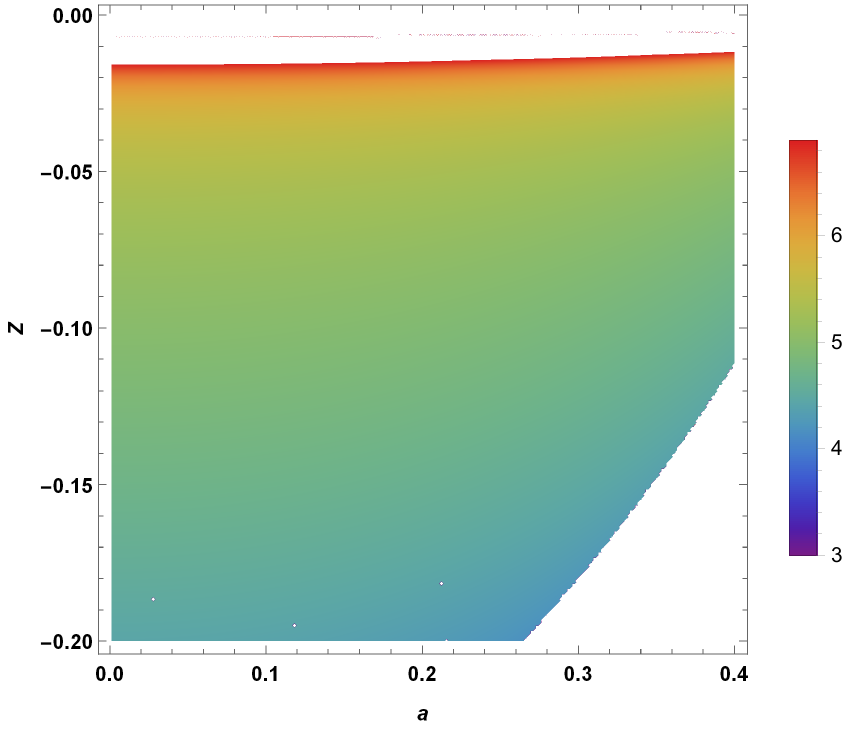}}
     \newline
\caption{ The shadow diameter as a function of $(a,Z)$ which shows acceptable regions in agreement with observational data of M87$^\star$. Here, we have set $Q=0.2 $ and $\mu=0.4 $ along with $M = 1$.}
\label{Fig10a}
\end{figure}

Now we use the mentioned observables to estimate the range of parameters with EHT data. Fig. \ref{Fig10a} shows 
the variation of the shadow diameter $ (d_{sh}) $ with the change in parameter $Z$ and the spin parameter $a$ for a fixed electric charge. The shaded area indicates the region where the resulting shadow is physically acceptable. However, it can be seen from this figure that the constraint (\ref{EqdM87b}) is not satisfied in any range of the parameter space $(a,Z)$. In other words, the shadow consistent with the M87* data cannot be observed in any range of $Z$ and $a$ for the BH with electric charge $Q=0.2$.  This reveals the fact that the electric charge plays a key role in achieving results consistent with the EHT data. Table \ref{tableI} verifies this claim and illustrates that the constraint (\ref{EqdM87b}) is satisfied for a limited range of $Q$, such that for a slowly rotating BH in dS space-time, $ d_{sh} $ is in agreement with
M87* data within $ 1\sigma $-error ($ 2\sigma $-error) for $ 0.247<Q<0.267 $ ($ 0.225<Q<0.247 $). While for BHs in  AdS space-time, a compatible result with  M87* data will not observed in any range of $ Q $ and $ a $.

\begin{figure}[!htb]
\centering
\subfloat[]{
        \includegraphics[width=0.325\textwidth]{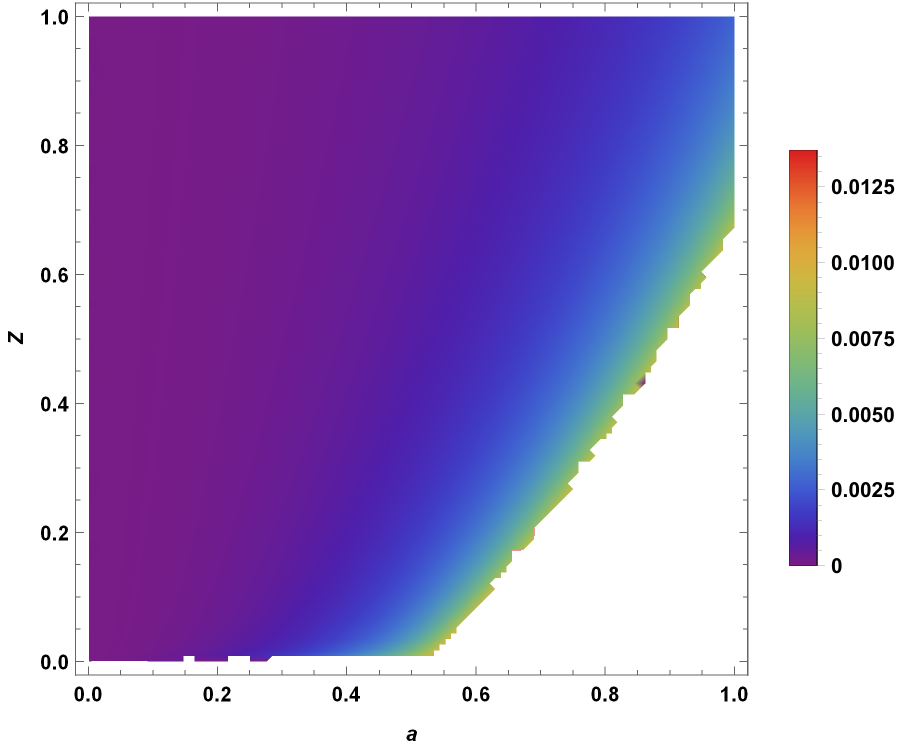}}
\subfloat[]{
     \includegraphics[width=0.31\textwidth]{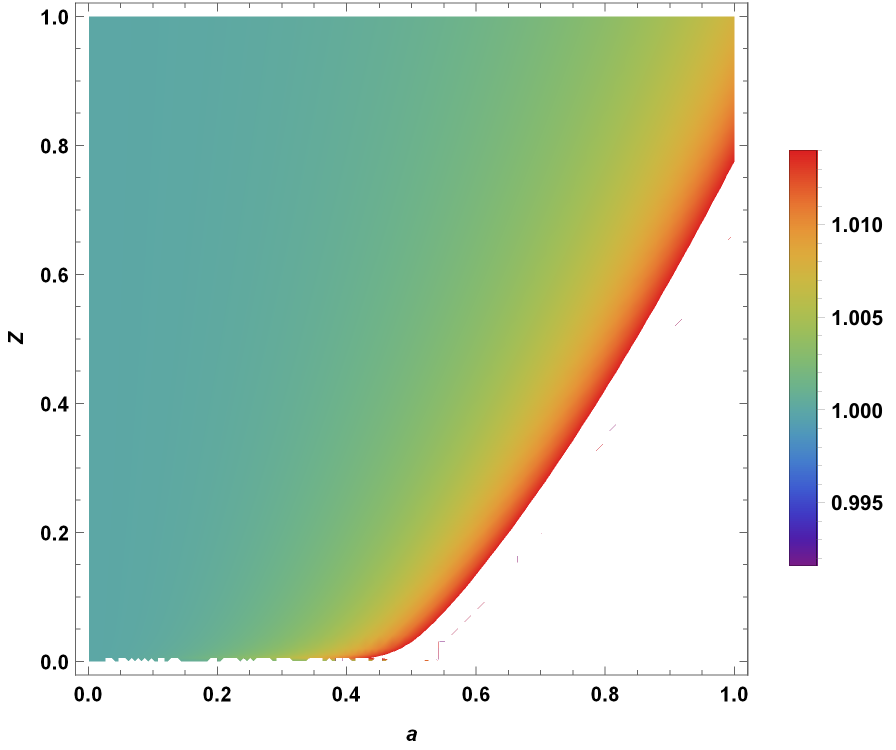}}
     \newline
\caption{The density plots of the circularity deviation $ \Delta C $ (left panel) and axial ratio $ D_{x} $ (right panel) in the $(a - Z)$ plane while keeping 
$Q=0.2$ and $\mu=0.4$ fixed.}
\label{Fig11}
\end{figure}

To restrict BH parameters using the EHT constraints on $ \Delta C $ and $ D_{x} $, we have plotted Fig. \ref{Fig11} which displays the variation of $ \Delta C $ and $ D_{x} $ in the parameters plane $(a,Z)$.  The shaded area shows the range of $Z$ and $a$ which satisfies the constraints $ \Delta C \lesssim 0.1 $ and $1 < D_{x} \lesssim 4/3$. According to our analysis, $ \Delta C \leq 0.01 $ for all parameter values, and $D_{x}$ lies in the range $1 < D_{x} < 1.014$ for all $a$ and $Z$. Therefore, the whole parameter space satisfies both constraints $ \Delta C \lesssim 0.1 $ and $1 < D_{x} \lesssim 4/3$. This reveals that it is difficult to constrain the parameters in $R^{2}$ gravity or to distinguish this modified gravity from GR using these two observables. 

\begin{figure}[!htb]
\centering
     \subfloat[]{
        \includegraphics[width=0.31\textwidth]{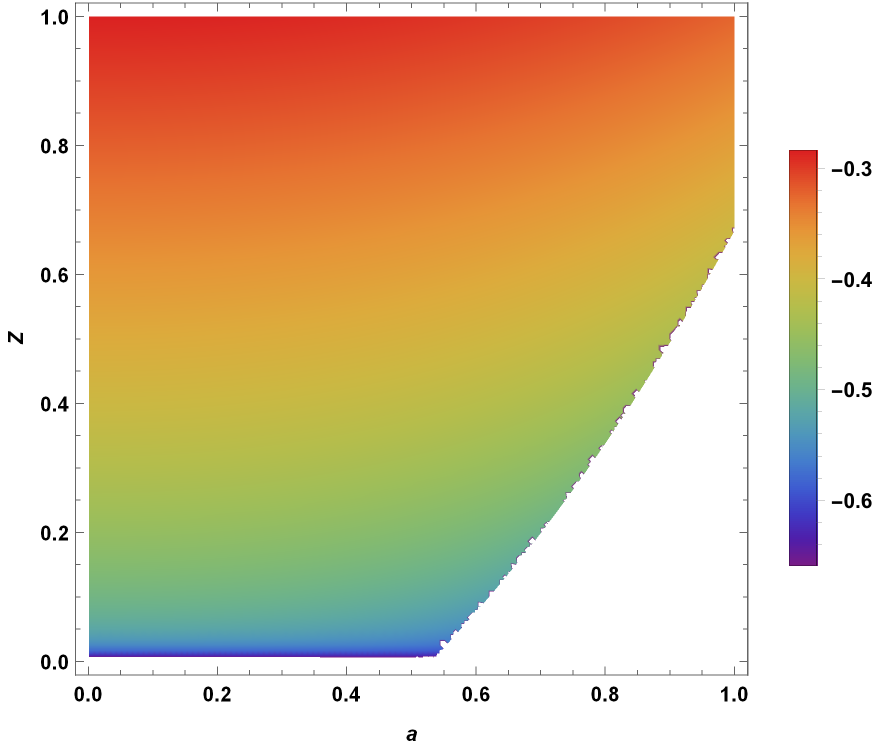}}
\subfloat[]{
     \includegraphics[width=0.32\textwidth]{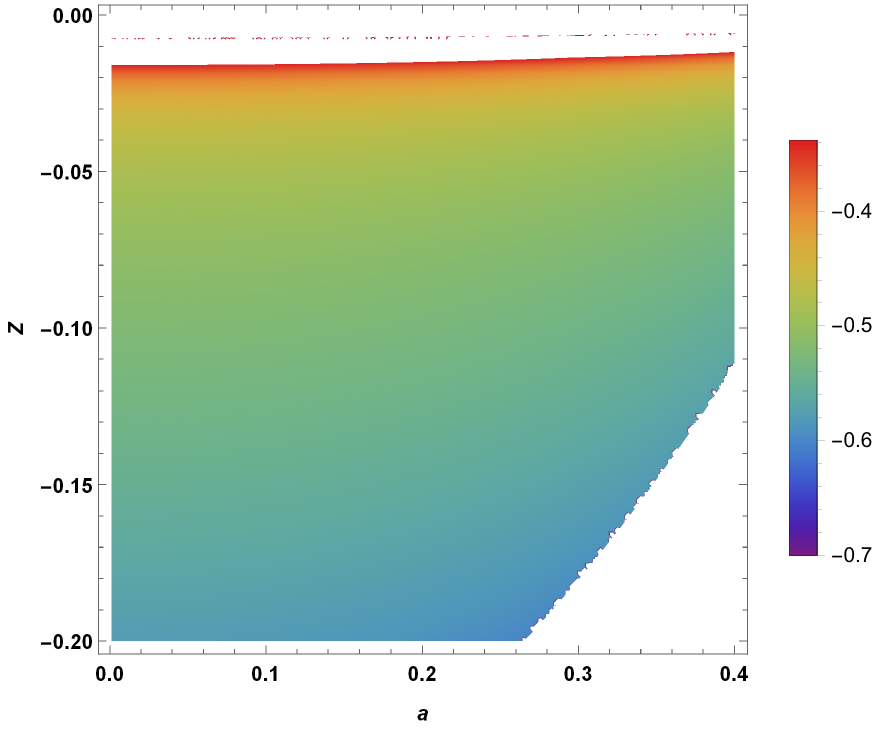}}
     \newline
\caption{The shadow diameter deviation $\delta $
from that of the Schwarzschild BH as a function of
$(a, Z)$ while keeping $ Q=0.2$ and $\mu=0.4$ fixed.}
\label{Fig10b}
\end{figure}

Fig. \ref{Fig10b} shows the deviation of the resulting shadow from the Schwarzschild shadow diameter, denoted by $\delta$, in $(a,Z)$ plane. In the unshaded area, the shadow size is imaginary which is non-physical. As was already mentioned, the EHT observations of M87* measured the bound on the Schwarzschild deviation as $ \delta=-0.01 \pm 0.17 $.  Fig. \ref{Fig10b} demonstrates that the mentioned constraints can be satisfied for some ranges of parameters. The exact range is addressed in Table \ref{tableII}, indicating that in dS spacetime and for the fixed electric charge, $\delta$ lies within $ 1\sigma $ uncertainty for the range $0<a<0.3$. While for a fixed rotation parameter, the range $0.234<Q<0.265$ satisfies $ 1\sigma $ bound. Regarding the AdS spacetime, it can be seen from this table that no range of parameters satisfies the $ 1\sigma $ bound. However, $ 2\sigma $ bound can be maintained for a limited range of $a$ and $Q$.

\subsection{Constraints from the EHT observations of Sgr A$^\star$}
We now test and constrain the black hole parameters using EHT observations from Sgr A$^\star$.  According to the recent results of EHT collaboration, the mass and the distance of Sgr A$^\star$ are
$ M=(4.3 \pm 0.013) \times 10^{6} M_{\odot} $ and $\mathbb{D}= 8277\pm 33 pc $, respectively. The angular diameter of the emission ring of Sgr A$^\star$ turns out to be $\theta_{d} = (51.8 \pm 2.3)\mu as $. However, the angular diameter of the shadow is $\theta_{sh} = (48.7 \pm 7)\mu as$. Using these reported numbers, the diameter of the shadow image for Sgr A$^\star$ is determined as \cite{Akiyama22L12,Akiyama22L17}
\begin{equation}
d_{Sgr. A^\star}\equiv \frac{\mathbb{D}\theta_{sh}}{M}\approx 9.5 \pm 1.4.
\label{EqdSgrb}
\end{equation}

To calculate Schwarzschild shadow deviation $ \delta $ based on
Sgr A$^\star$ observations, the EHT employed the two separate priors for the
shadow size from the Keck and  VLTI observations to estimate the bounds on $ \delta $ as follows \cite{Akiyama22L12,Akiyama22L17}
\begin{equation}
\delta=\left\{
  \begin{array}{ll}
  -0.08^{+0.09}_{-0.09}  & \;\;(\mbox{VLTI}) \\
  -0.04^{+0.09}_{-0.10}  & \;\;(\mbox{Keck})
  \end{array}
\right..
\label{keck}
\end{equation}

\begin{figure}[!htb]
\centering
\subfloat[]{
        \includegraphics[width=0.31\textwidth]{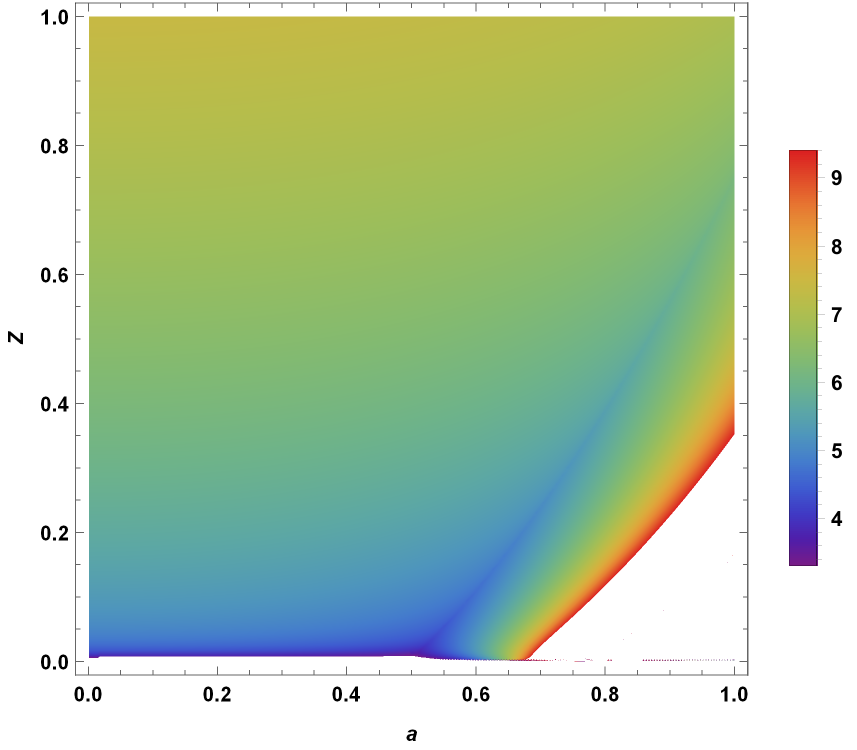}}
\subfloat[]{
     \includegraphics[width=0.32\textwidth]{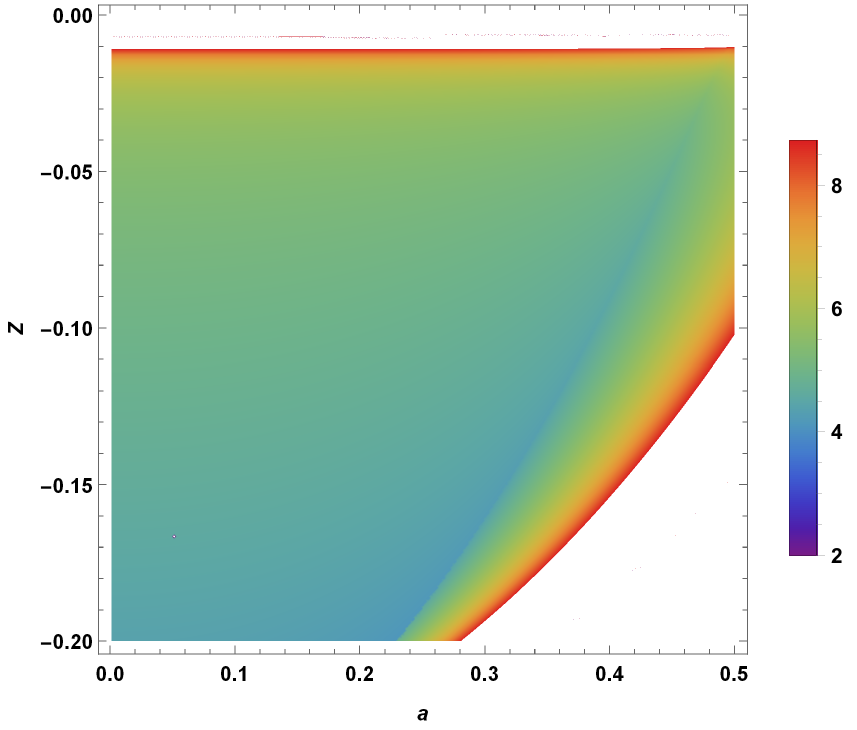}}
     \newline
     \subfloat[]{
        \includegraphics[width=0.31\textwidth]{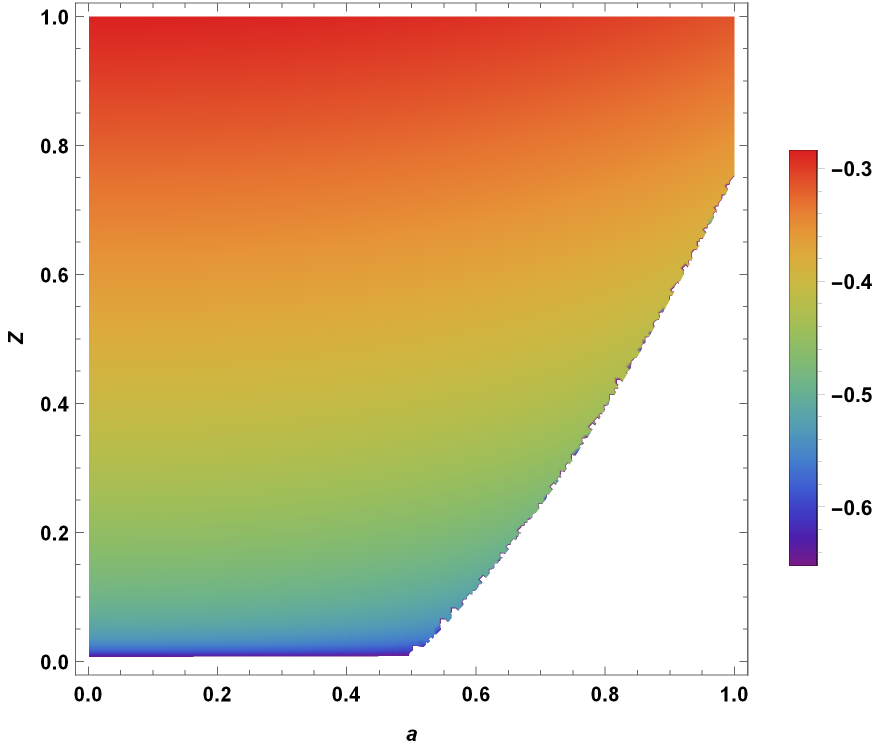}}
\subfloat[]{
     \includegraphics[width=0.32\textwidth]{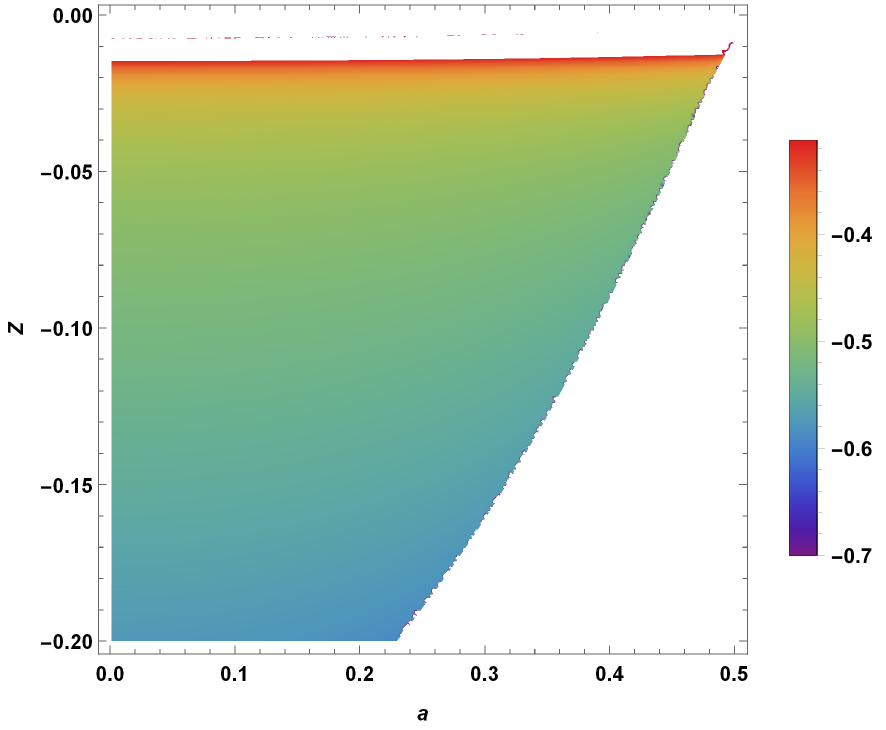}}
     \newline
\caption{The shadow diameter as a function of $(a,Z)$ showing acceptable regions in agreement with observational data from Sgr A$^\star$ (Top row). The shadow diameter deviation $ \delta$
from that of the Schwarzschild BH as a function of
$(a, Z)$ (Bottom row). Here, we have set $ Q=0.2 $ and $ \mu=0.4$ along with $M = 1$.}
\label{Fig12}
\end{figure}

To find the allowed region of parameters using EHT data of Sgr A$^\star$, we used the constraint related to two observables, the shadow
size (\ref{EqdSgrb}) and the fractional deviation (\ref{keck}) and plotted Fig.~\ref{Fig12}. The top panels of Fig.~\ref{Fig12} display the variation of the shadow diameter $d_{sh}$ with the change of $Z$ and $a$. It is clear that the constraint (\ref{EqdSgrb}) is satisfied for some ranges of parameters in both dS and AdS spacetime. The exact range is addressed in Table \ref{tableI}. According to this table, in dS spacetime, only slowly rotating BHs have a compatible shadow with Sgr A$^\star$ data. While in AdS spacetime, fast-rotating BHs provide a result consistent with Sgr A$^\star$ data. The bottom panels of this figure illustrate the deviation $\delta$ in $(a,Z)$ plane. It is clear from these diagrams that the constraint (\ref{keck}) is not satisfied in any range of the rotation parameter and $Z$ for a fixed electric charge. However, Table \ref{tableII} shows that the mentioned constraint will be satisfied for a very limited region of electric charge in dS spacetime. While in AdS spacetime, no range of parameters $a$ and $Q$ can satisfy the Keck and VLTI bound. It should be noted that the unshaded area indicates where the resulting shadow has an imaginary value.

From Table \ref{tableI}, one can find that the resulting shadow of KN-AdS BHs in $ R^{2} $ gravity is not in agreement with data of M87$^\star$. Hence, Sgr A$^\star$ black hole can be a
suitable model for such BHs. Regarding the KN-dS BHs, as we see from the left column of this table, the resulting shadow diameter is consistent with EHT data of
M87$^\star$/Sgr A$^\star$ within $1\sigma/2\sigma$ uncertainty. Therefore, comparing to M87$^\star$,  observations of Sgr A$^\star$ can provide more precise constraints on the BH parameters in $R^{2}$ gravity.

Table \ref{tableII} shows the constraints on BH parameters using the fraction deviation $ \delta $. From this Table, none of the parameters of KN-AdS BHs can satisfy Keck and VLTI bounds, while for KN-dS BHs, Keck/VLTI bounds can put constraints on the parameters. As a result, the KN-dS BH is a suitable candidate for M87$^\star$/Sgr A$^\star$ supermassive BHs.

\begin{table*}[t]
\caption{\label{tab1} Constraints on BH parameters using the diameter of the shadow image of M87$^\star$ and Sgr A$^\star$.}
      \centering
      \begin{tabular}{|c| |c|c| |c|c||}
\hline
  &\multicolumn{2}{c||}{KN-dS in $ R^{2} $ gravity} & \multicolumn{2}{c||}{KN-AdS in $ R^{2} $ gravity} \\
\cline{2-5}
Data & \multicolumn{2}{c||}{ $  Q=0.25$ and $ Z=-0.015 $} & \multicolumn{2}{c||}{$  Q=0.25$ and $ Z=0.2$} \\
\cline{2-5}
 & \multicolumn{1}{l|}{\quad\quad \quad\quad 1$\sigma$ }            & \quad\quad 2$\sigma$ \quad\quad           & \multicolumn{1}{l|}{\quad\quad \quad 1$\sigma$ }            & \quad\quad 2$\sigma$ \quad\quad           \\ \hline
M87*                & \multicolumn{1}{l|}{\quad\quad $a\in (0, 0.17)$} & $a\in [0.17, 0.365]$ & \multicolumn{1}{l|}{\quad\quad \quad$--$} & $--$ \\ \hline
Sgr A$^\star$                & \multicolumn{1}{l|}{\quad\quad$a\in (0, 0.58]$} & $a\in (0.58, 0.62]$ & \multicolumn{1}{l|}{$a\in [0.848, 0.905]$} & $a\in [0.795, 0.848)\cup (0.905, 0.924]$ \\ \hline \hline
 Data& \multicolumn{2}{c||}{$  a=0.1$ and $ Q=0.25 $} & \multicolumn{2}{c||}{$  a=0.85$ and $ Q=0.25 $} \\
\cline{2-5}
 & \multicolumn{1}{l|}{\quad\quad \quad\quad 1$\sigma$ }            & \quad\quad 2$\sigma$ \quad\quad           & \multicolumn{1}{l|}{\quad\quad \quad 1$\sigma$ }            & \quad\quad 2$\sigma$ \quad\quad           \\ \hline
M87*                 & \multicolumn{1}{l|}{$Z\in[-0.015,-0.013)$} & $Z\in[-0.018,-0.015)$ & \multicolumn{1}{l|}{\quad\quad \quad$--$} & $--$ \\ \hline
Sgr A$^\star$                & \multicolumn{1}{l|}{$Z\in[-0.018,-0.014]$} & $Z\in(-0.014,-0.013]\cup [-0.025,-0.018)$ & \multicolumn{1}{l|}{$Z\in [0.143,0.202)$} & $Z\in [0.128, 0.143)\cup [0.202, 0.285]$ \\ \hline \hline
Data& \multicolumn{2}{c||}{$  a=0.1$ and $ Z=-0.015 $} & \multicolumn{2}{c||}{$  a=0.85$ and $ Z=0.2 $} \\
\cline{2-5}
 & \multicolumn{1}{l|}{\quad\quad \quad\quad 1$\sigma$}            & \quad\quad 2$\sigma$ \quad\quad           & \multicolumn{1}{l|}{\quad\quad \quad 1$\sigma$ }            & \quad\quad 2$\sigma$ \quad\quad           \\ \hline
M87*                & \multicolumn{1}{l|}{\quad$Q\in(0.247,0.267]$} & $Q\in(0.225,0.247]$ & \multicolumn{1}{l|}{\quad\quad \quad$--$} & $--$ \\ \hline
Sgr A$^\star$                & \multicolumn{1}{l|}{\quad$Q\in [0.226,0.257]$} & $Q\in[0.187,0.226)\cup (0.257,0.265]$ & \multicolumn{1}{l|}{\quad$Q\in [0, 0.3]$} & $Q\in (0.3, 0.93]$ \\ \hline 
\end{tabular}
\vspace{1ex}
\label{tableI}
\end{table*}

\begin{table*}[t]
\caption{\label{tab2} Constraints on BH parameters using the fraction deviation observable $ \delta $.}
      \centering
      \begin{tabular}{|c| |c|c| |c|c||}
\hline
  &\multicolumn{2}{c||}{KN-dS in $ R^{2} $ gravity} & \multicolumn{2}{c||}{KN-AdS in $ R^{2} $ gravity} \\
\cline{2-5}
Data & \multicolumn{2}{c||}{ $  Q=0.25$ and $ Z=-0.015 $} & \multicolumn{2}{c||}{$  Q=0.25$ and $ Z=0.8$} \\
\cline{2-5}
 & \multicolumn{1}{l|}{\quad\quad\quad\quad 1$\sigma$ }            & \quad\quad 2$\sigma$ \quad\quad           & \multicolumn{1}{l|}{\quad\quad 1$\sigma$ \quad\quad}            & \quad\quad 2$\sigma$ \quad\quad           \\ \hline
M87*                & \multicolumn{1}{l|}{\quad\quad\quad$a\in (0, 0.3]$} & $a\in (0.3, 0.48]$ & \multicolumn{1}{l|}{\quad\quad$--$} & $a\in (0, 0.82]$ \\ \hline
Sgr A$^\star$ (VLTI)               & \multicolumn{1}{l|}{\quad\quad\quad$a\in (0, 0.49)$} & $--$ & \multicolumn{1}{l|}{\quad\quad$--$} & $--$ \\ \hline
Sgr A$^\star$ (Keck)                & \multicolumn{1}{l|}{\quad\quad\quad$a\in (0, 0.46]$} & $a\in (0.46, 0.49)$ & \multicolumn{1}{l|}{\quad\quad$--$} & $--$ \\ \hline \hline
 Data& \multicolumn{2}{c||}{$  a=0.1$ and $ Q=0.25 $} & \multicolumn{2}{c||}{$  a=0.5$ and $ Q=0.25 $} \\
\cline{2-5}
 & \multicolumn{1}{l|}{\quad\quad\quad\quad 1$\sigma$ }            & \quad\quad 2$\sigma$ \quad\quad           & \multicolumn{1}{l|}{\quad\quad 1$\sigma$ \quad\quad}            & \quad\quad 2$\sigma$ \quad\quad           \\ \hline
M87*                 & \multicolumn{1}{l|}{$Z\in[-0.017,-0.013]$} & $Z\in[-0.024,-0.017)$ & \multicolumn{1}{l|}{\quad\quad$--$} & $Z\in[0.71,1)$ \\ \hline
Sgr A$^\star$ (VLTI)                & \multicolumn{1}{l|}{$Z\in[-0.0169,-0.0144)$} & $Z\in[-0.0195,-0.0169)$ & \multicolumn{1}{l|}{\quad\quad$--$} & $--$ \\ \hline
Sgr A$^\star$ (Keck)               & \multicolumn{1}{l|}{$Z\in[-0.0163,-0.0142]$} & $Z\in(-0.0189,-0.0163)$ & \multicolumn{1}{l|}{\quad\quad$--$} & $--$ \\ \hline \hline
Data& \multicolumn{2}{c||}{$  a=0.1$ and $ Z=-0.015 $} & \multicolumn{2}{c||}{$  a=0.5$ and $ Z=0.8 $} \\
\cline{2-5}
 & \multicolumn{1}{l|}{\quad\quad\quad\quad 1$\sigma$ }            & \quad\quad 2$\sigma$ \quad\quad           & \multicolumn{1}{l|}{\quad\quad 1$\sigma$ \quad\quad}            & \quad\quad 2$\sigma$ \quad\quad           \\ \hline
M87*                & \multicolumn{1}{l|}{\quad$Q\in(0.234,0.265]$} & $Q\in[0.191,0.234]$ & \multicolumn{1}{l|}{\quad\quad$--$} & $Q\in[0,0.475)$ \\ \hline
Sgr A$^\star$ (VLTI)                 & \multicolumn{1}{l|}{\quad$Q\in(0.235,0.255]$} & $Q\in[0.218,0.235]$ & \multicolumn{1}{l|}{\quad\quad$--$} & $--$ \\ \hline
Sgr A$^\star$ (Keck)               & \multicolumn{1}{l|}{\quad$Q\in (0.233,0.258)$} & $Q\in[0.258,0.263]$ & \multicolumn{1}{l|}{\quad\quad$--$} & $--$ \\ \hline 
\end{tabular}
\vspace{1ex}
\label{tableII}
\end{table*}

\section{\label{Sec:conclusion} Conclusions}

In this paper, we considered a charged AdS/dS BH and the rotating counterpart of the static solutions in $R^{2}$ gravity and investigated their optical features, including the geometrical shape of the shadow and the energy emission rate. It is important to emphasize that, in order to have confidence in the conclusions drawn from the observations, it is crucial to analyse the effects of $R^2$ gravity on null geodesics and the energy emission rate to provide valuable insights into BH properties within different gravity models. We began to examine the allowed regions of the parameters, where a physical BH solution can exist and showed the admissible parameter space plot ($Q-Z$) around the charged AdS/dS BHs for various possible cases. We observed that the admissible parameter space region decreases under the influence of the non-zero cosmological constant $\mu$ for both positive and negative cases of the coupling parameter $Z$ in $R^2$ gravity, as depicted in Fig.~\ref{Fig1}. 

We then studied the BH shadow under the combined effects of the BH charge $Q$, non-zero cosmological constant $\mu$ and coupling parameter $Z$ within the $R^2$ gravity. Further, we showed the BH shadow size effectively decreases due to the impact of $Q$, leading to the shape of shadow sphere shifting towards smaller $r_{sh}$. Unlike $Q$, it was observed from the results that the BH shadow size increases under the influence of positive $Z>0$. Interestingly, we observed that the opposite is true for negative case of $Z<0$, similarly to what is observed in the BH shadow size due to $Q$. To provide upper constraints on the BH parameters in $R^2$ gravity, we applied the recent EHT observations of the supermassive BH at the center of M87$^{\star}$ to the obtained theoretical results. We showed the constraint space of $(Q,Z)$ in $R^{2}$ gravity within the confidence levels of $1\sigma$ and $2\sigma$ using the EHT observations of M87$^{\star}$. We observed that the positive $Z>0$ and $Q$ support each other, causing the upper value of $Z$ to increase with an increasing upper value of $Q$. We noted that the opposite result occurs for the negative case of $Z$, leading to its upper limits that decrease with an increasing upper limit of the BH charge $Q$. 

Besides, we studied the shadow diameter deviation from a Schwarzschild BH and found out that $\delta$ approaches positive values as the parameter space of $(Q,Z)$ increases, meaning that a charged AdS BH in $R^{2}$ gravity with the same mass as Schwarzschild, has a greater (smaller) shadow size than that of the Schwarzschild BH for large (small) values of the parameters $Q$ and $Z$. While for charged dS BHs, the shadow size gets greater than the Schwarzschild shadow for small values of the parameters. Additionally, we studied the behavior of energy emission rate as a function of $\omega$ around the charged AdS/dS BH in $R^2$ gravity. We noticed that the evaporation process will be faster for a BH located in a weak electric field, illustrating that neutral BHs have shorter lifetimes compared to their charged counterparts in $R^{2}$ gravity. Our analysis also showed that the corresponding BHs evaporate very slowly/fast in AdS/dS background which reveals the fact that a black hole has a longer/shorter lifetime in AdS/dS spacetime. 

Further, we generalized the static and spherically symmetric charged BH solution in $R^{2}$ gravity to the Kerr-like rotating BH solution by employing the Newman-Janis algorithm (NJA). We then analyzed the allowed regions of the parameters for a valid physical BH solution and demonstrated the admissible parameter space plot ($a-Z$). We showed that the allowed region of the admissible parameter space ($a-Z$) shrinks due to the influence of 
$Q$ for both positive and negative cases of the coupling parameter $Z$. 

Also, we delved into the impacts of $Q$, $Z$ and the spin parameter $a$ on the rotating charged BH shadow boundary/size within the theory of $R^2$ gravity. It was shown that the size of the shadow boundary becomes smaller due to the impact of $Q$, leading to the shape of the shadow sphere shrinking with an increasing $Q$. It does however increase as a consequence of the influence of $Z$, leading to the shadow sphere becoming possibly larger, as shown in Fig.~\ref{Fig8}. Note that the situation turns out to be different in the inclusion of the spin parameter $a$. The symmetric form of the BH shadow, as expected, breaks down as the horizontal diameter becomes smaller than the vertical diameter with varying spin parameter $a$. A similar effect was also observed for varying values of the inclination angle $\theta$. This asymmetric feature does however disappear in the limit of $\theta\to 0$ for a fixed $a$, as well as of $a\to 0$ for a fixed $\theta$. Finally, we examined the behavior of the energy emission rate around the rotating charged BH. According to our findings, the rotation parameter decreases the energy emission rate, while the electric charge has an increasing effect on the emission rate. This reveals the fact that fast-rotating BHs or black holes located in a weak electric field have longer lifetime. Regarding the influence of the coupling parameter $Z$ on the energy emission, we figured out that increasing this parameter leads to increasing energy emission in both AdS and dS spacetime. In other words, the black hole has a shorter lifetime when the effect of $R^2$ gravity gets stronger.

Finally, we considered the obtained rotating BH as a supermassive BH and used precise measurements from EHT observations for M87$^{\star}$ and Sgr A$^\star$ and compared these observational data with theoretical results to obtain the best-fit constraints on the BH parameters in $R^2$ gravity, as shown in Tables~\ref{tab1} and \ref{tab2}. Based on the results, we observed that for KN-AdS BHs in $ R^{2} $ gravity,  the resulting shadow is not in agreement with data of M87$^\star$. While observational data of Sgr A$^\star$ can impose constraints on parameters. This shows that Sgr A$^\star$ can be a suitable model for such BHs. Regarding the KN-dS BHs, our results indicated that the resulting shadow is consistent with EHT data of M87$^\star$/Sgr A$^\star$ within $1\sigma/2\sigma$ uncertainty. However, comparing to M87$^\star$,  observations of Sgr A$^\star$ can provide more precise constraints on the BH parameters. Studying the fraction deviation $\delta$ illustrated that none of the parameters of KN-AdS BHs can satisfy Keck and VLTI bounds, while for KN-dS BHs, Keck/VLTI bounds can put constraints on the parameters. Taken together, we found that the KN-dS BH is a suitable candidate for the supermassive BHs at the center of the host galaxies M87$^\star$ and Sgr A$^\star$. 

Given the importance of BH shadows and their observational signatures, these results hold significant astrophysical importance, as they do not rule out the influence of $R^2$ gravity. Our theoretical findings may thus offer insights into the nature of $R^2$ gravity and help constrain alternative models to GR when interpreting astrophysical observations and predictions.

\section*{ACKNOWLEDGEMENT}

The research is supported by the National Natural Science Foundation of China under Grant No. W2433018.

\bibliography{Ref}
\end{document}